\documentclass[10pt,aps,twocolumn,prc,superscriptaddress,noshowpacs,nofootinbib,noshowkeys,floatfix]{revtex4}
\usepackage[dvips]{graphics,graphicx}
\usepackage[colorlinks=true,linktocpage=true,linkcolor=blue,citecolor=blue,urlcolor=blue]{hyperref}
\usepackage[usenames,dvipsnames]{color}
\usepackage{amsmath, amssymb}
\usepackage{multirow}
\usepackage{longtable}
\usepackage{bm}
\usepackage{color}
\usepackage{xcolor}
\usepackage{pbox}
\usepackage[normalem]{ulem}  

\renewcommand\sout{\bgroup \color{blue} \ULdepth=-.5ex \ULset}
\usepackage{lipsum}
\usepackage{lipsum}
\usepackage{color,soul}
\setul{0.5ex}{0.3ex}
\definecolor{Red}{rgb}{1,0.0,0.0}
\setulcolor{Red}
\usepackage{hyperref}

\usepackage{makecell}

\setstcolor{red}

\begin{document}
	
	\preprint{}
	
	\title{Examination of thermalization of quarkonia at energies available at the CERN Large Hadron Collider}
	
	\author{Deekshit Kumar}
	\email{deekshitkumarvecc@gmail.com}
	\affiliation{Homi Bhabha National Institute, Anushakti Nagar, Mumbai 400094, India}
	
	\author{Nachiketa Sarkar}
	\email{nachiketa@niser.ac.in}
	\affiliation{School of Physical Sciences, National Institute of Science Education and Research, An OCC of Homi Bhabha National Institute, Jatni-752050, India} 
	
	\author{Partha Pratim Bhaduri}
	\email{partha.bhaduri@vecc.gov.in}
	\affiliation{Homi Bhabha National Institute, Anushakti Nagar, Mumbai 400094, India}
	\affiliation{Variable Energy Cyclotron Centre, 1/AF Bidhan Nagar, Kolkata 700 064, India}
	
	\author{Amaresh Jaiswal}
	\email{a.jaiswal@niser.ac.in}
	\affiliation{School of Physical Sciences, National Institute of Science Education and Research, An OCC of Homi Bhabha National Institute, Jatni-752050, India} 
	
	\date{\today}
	
	\begin{abstract}
		We analyze the relative yields of different bottomonia and charmonia states produced in Pb-Pb, p-Pb and high multiplicity p-p collisions at LHC, within a semi-classical grand canonical ensemble approach. The underlying assumption is the early thermalization and subsequent freezeout of these heavy hadrons resulting in their chemical freezeout at a temperature of approximately $230$~MeV, significantly higher than that of light and strange hadrons. The systematic dependence of the freezeout temperature on the collision centrality is also investigated in details.
	\end{abstract}
	
	\maketitle
	
	
	\section{Introduction}
	
	Quarkonium are the bound states of a heavy Quark ($Q$) and its anti-quark ($\bar{Q}$). The bound state of a charm (bottom) quark and its anti-quark is called charmonium (bottomonium). The binding is believed to be hindered in presence of a thermally deconfined medium, in a process analogous to the Debye screening in an electromagnetic plasma~\cite{Matsui:1986dk}. Since a thermalized partonic medium is likely to be formed in relativistic heavy-ion collisions, but not concretely established in proton-proton (p-p) and proton-nucleus (p-A) collisions, the production yield of the quarkonium states is expected to be systematically different in two cases. Suppression of heavy quarkonium states is believed to be a promising diagnostic probe of the properties of the hot and dense medium created in high-energy heavy-ion collisions~\cite{VOGT1999197, Satz:2005hx, Kluberg:2009wc, Rapp:2008tf, Zhao:2020jqu}. 
	
	If quark-gluon plasma (QGP) is formed in the collision zone, the confining potential of heavy quark-antiquark pairs is expected to be screened by the color charges leading to the in-medium modification of the quarkonium spectral functions. In addition, the medium-induced imaginary potential of the quarkonium system dubbed in terms of the inelastic scattering with hard partons leads to the thermal broadening of their in-medium width, at any finite temperature. These in-medium effects have been intensely studied based on lattice quantum chromodynamics (QCD) and effective field theories of QCD~\cite{Mocsy:2013syh, Rothkopf:2019ipj}. The resulting dissociation (or melting) of the quarkonium states in a hot plasma depends on the temperature, with loosely bound states melting earlier than the strongly bound ones. Quarkonium dissociation is thus expected to occur sequentially, reflecting the increasing values of their binding energies~\cite{Karsch:2005nk, Bhaduri:2008zs}. This leads to the characteristic sequential suppression pattern of different quarkonium states in heavy ion collisions, compared to their production in p-p collisions.
	
	Experimentally charmonium production in heavy-ion collisions was measured at SPS~\cite{NA50:2000brc, NA60:2007ewx}, RHIC~\cite{PHENIX:2006gsi, PHENIX:2011img, STAR:2019fge} and LHC~\cite{ALICE:2013osk, ALICE:2015jrl, ALICE:2016flj,CMS:2012bms, ALICE:2019lga, ATLAS:2018hqe, CMS:2016wgo, CMS:2017uuv, ALICE:2022jeh}. The systematic difference in the production pattern between nucleus-nucleus (A-A) collisions and p-p collisions is quantified by the nuclear modification factor ($R_{AA}$), defined as the ratio between the quarkonium yield in nuclear collisions and the yield in p-p collisions scaled by the number of binary nucleon-nucleon collisions. At the LHC, suppression of $J/\psi$ production in nuclear collisions is weaker than the measurements at lower energies. In central collisions at LHC, the $R_{AA}$ is much larger than at RHIC energy for low-$p_{T}$ $J/\psi$. The $p_{T}$ dependence of $R_{AA}$ is also significantly different at RHIC and LHC. At RHIC, the $p_T$ dependence is rather flat, but a clear decreasing trend with increasing $p_{T}$ is observed at LHC. Comparison of the $p_{T}$ dependence of the inclusive $J/\psi$ $R_{AA}$ in various rapidity windows, also exhibits the decreasing trend with more suppression at higher rapidities. These observations are commonly attributed to the significant regeneration of the charmonium states at low $p_{T}$, from the initially deconfined and uncorrelated $c,\overline{c}$ pairs. Regeneration is more important at higher energies due to the larger amount of $c\overline{c}$ pairs present in the medium.
	
	In the bottomonium sector, measurements at the LHC are pioneered by the CMS Collaboration, with the observation of strong suppression of $\Upsilon(1S)$ state in Pb-Pb collisions~\cite{CMS:2012gvv}. The production of the excited $\Upsilon(2S)$ state suffered a much more stronger suppression. The $b\overline{b}$ production cross section is much lower than that of $c\overline{c}$, owing to their large mass. Bottonium states are thus expected to be much less affected by the regeneration effect~\cite{Du:2017qkv}. Attempts were also made to measure the nuclear modification factors in proton-nucleus (p-A) collisions at LHC, where formation of a deconfined plasma phase was not expected prior to the LHC measurements. Data indicate a significant modification of the $\Upsilon(1S)$ production at mid-rapidity and forward rapidity, in accordance with the expectations from the initial state effects. However a stronger suppression is observed at backward rapidity, unaccountable by the known initial state effects. Additionally, excited states were seen to undergo stronger suppression, analogous to the pattern observed in A-A collisions. While $R_{AA}$ is an useful observable to look for signals of medium formation in nuclear collisions, it is not the most appropriate quantity to study thermal physics. In literature, more popular choices to address the issue of thermalization includes the measurement transverse momentum ($p_T$) spectra and the elliptic flow ($v_2$), the second Fourier coefficient of the azimuthal anisotropy.
	In literature, the measured $p_T$ spectra are analyzed within hydrodynamics inspired blast wave model to address the issue of the degree to which heavy quarks thermalize in the fireball and inherit the collective expansion of the medium. In Ref.~\cite{Gazdzicki:1999rk}, the possible thermal features of charmonium production was first pointed out by analyzing the $p_T$ spectra of $J/\psi$ (and $\psi'$) mesons in $\sqrt{s_{NN}}=17.3$ GeV Pb-Pb collisions at SPS. Subsequently the $p_{T}$ distribution of $J/\psi$ mesons measured at RHIC~\cite{Bugaev:2002fd} and LHC~\cite{Andronic:2019wva} have also been analyzed within such boosted thermal model framework. The underlying assumption in such analyses is the coincidence of the kinetic freeze-out temperature with that of chemical freeze-out due to negligible rescattering cross sections of charmonia in the hadronic phase. For bottomonia, $\Upsilon$ $p_{T}$ spectra are currently available from CMS collaboration in $\sqrt{s_{NN}}=2.76$ TeV Pb-Pb collisions, for $0-100 \%$ centrality~\cite{CMS:2016rpc}. Whether these spectra come in agreement with a blast wave model description is difficult to judge~\cite{Reygers:2019aul}. Recent measurements by both CMS and ALICE collaborations in $\sqrt{s_{NN}}=5.02$ TeV Pb-Pb collisions at LHC indicate a smaller $v_2$ of $\Upsilon(1S)$ states~\cite{ALICE:2019pox, CMS:2019uhg} than the one measured for inclusive $J/\psi$~\cite{ALICE:2017quq, ALICE:2018bdo, ALICE:2020pvw} mesons, by ALICE collaborations. The measured $v_{2}(p_{T})$ of the inclusive $J/\psi$ mesons can be explained upto large $p_T$, using transport model incorporating the state-of-the-art $c$-quark phase space distributions~\cite{He:2021zej}, into the regeneration process of charmonia. On the other hand, the $v_{2}$ of the $\Upsilon$ mesons was found to be consistent with zero over the measured $p_{T}$ range. Within the associated uncertainty, the data can be explained both within models~\cite{Du:2017qkv,Bhaduri:2018iwr,Bhaduri:2020lur} which do not consider the $b$ quark thermalization (and hence negligible regeneration) in the fireball or within blast wave model framework~\cite{Reygers:2019aul} that assumes a scenario in which $b$ quarks thermalize and the $\Upsilon$ mesons inherit the collective expansion of the medium. Hence neither of the possibilities can be ruled out. 
	
	In Ref.~\cite{Gupta:2014ova} the authors have addressed the issue of quarkonium thermalization at LHC by analyzing the relative yields of different bottomonium states in $\sqrt{s_{NN}} = 2.76$ TeV Pb-Pb collisions, using then available data. The observed sequential suppression of the $\Upsilon$ family of mesons as measured by the CMS collaboration, was interpreted as the system being in thermal equilibrium in the fireball before freezing out. All the $Q\bar{Q}$ pairs present in the fireball are assumed to be produced in the initial hard scattering and eventually thermalize with the medium due to interaction. The production of heavy quark pairs is negligible at any conceivable plasma temperature. A simple dimensional analysis involving multiple scales including heavy quark mass $M_{Q}$ and temperature $T$ indicates that the net production rate of quarkonia at thermal equilibrium may be obtained by studying the spectral density of quarkonium states in thermal equilibrium. Lattice QCD simulations indicate increase in the decay width of the quarkonium states, $\Gamma$, at any finite $T$. The width is interpreted as the inclusive rate of all unbinding reactions in which quarkonium state disintegrates into unbound $Q\bar{Q}$ pairs. Moreover, at LHC energies, one also needs to account for the reverse recombination rate that can be estimated from the dissociation process via principle of detailed balance. 
	
	Note that large values of $\Gamma(T)$ would imply different bound and unbound $Q\bar{Q}$ states to reach thermal equilibrium with each other over a very short time. Production dynamics of the quarkonium states in nuclear collisions, where  a thermalized partonic medium is anticipated, is thus entirely different than that in elementary p-p collisions. Different quarkonium states would continue to be in thermal equilibrium, until $\Gamma$ falls below the expansion rate of the fireball and these heavy mesonic systems freeze out. This adopted picture of quarkonia production and evolution is similar in spirit with the statistical hadronization model (SHM) of quarkonia production. In this picture all the $Q\bar{Q}$ pairs are assumed to be produced in primary hard collisions and their total number remains conserved until hadronization. The number of $Q\bar{Q}$ pairs is thus effectively decoupled from the thermal description of the heavy quark hadronization at the QCD phase boundary, where charmonia (bottomonia) are formed from charm (bottom) quarks according to statistical weights evaluated at the chemical freeze-out . Apart from this non-thermal scale of the total number of charm quarks, the other parameters of the statistical hadronization scheme are the thermal parameters in a grand canonical description of the fireball. They are well constrained  by the chemical  freeze-out analysis of the light hadrons. Thus in SHM, the quarkonia are believed to have a chemical freeze-out temperature of $T_{f} \simeq 156$ MeV, same as all other hadrons. The so-called SHM for charmed hadrons (SHMC) provides a reasonable description of charmonium production~\cite{Andronic:2006ky,Andronic:2007bi,Andronic:2018vqh} in nuclear collisions. The same model can also describe the relative production of $\Upsilon(1S)$ and $\Upsilon(2S)$ mesons in $\sqrt{s_{NN}}=2.76$ TeV Pb-Pb collisions~\cite{Andronic:2017pug}. However in contrast to SHM, in the present model, the heavy quarkonia states are not made to undergo chemical freeze out at same temperature as for light hadrons. Instead the chemical freeze out of these heavy mesons is fixed by the available data, from the relative yield of different quarkonium states in Pb-Pb collisions. 
	
	Assuming all the quarkonium states to be in complete equilibrium with the fireball, their equilibrium densities are obtained in terms of the quarkonium mass, $M$ and temperature $T$ (for $M >> T$). The ratio $r[\Upsilon(nS)]$ of the yield of different excited bottomonium states to that of ground state is then constructed and contrasted with the available data. Assuming all the resonances in a given quarkonium family are freezing out together inside a fireball of nearly unit fugacity, $r$ contains only single parameter, the freeze out temperature, $T_{F}$. Accounting for the modification of primordial densities due to feed down decay from excited states and neglecting the errors in the branching ratio, the yield ratio of $\Upsilon(2S)$ and  $\Upsilon(1S)$ states indicated a preliminary estimation of the chemical freeze-out temperature $ T_{F} = 222_{-29}^{+28}$ MeV, which is much higher than that of light hadrons. One may note that the idea of quark flavour dependent sequential freeze out of the hadrons measured in relativistic heavy-ion collisions was first argued in Ref.~\cite{Chatterjee:2013yga}. A systematic analysis of the hadron yields within a thermodynamically consistent hadron  resonance  gas model at different collision energies, indicated the strange hadrons freezing out earlier at a higher temperature than the non strange light hadrons. Ref.~\cite{Gupta:2014ova} was the first extension of this two chemical freeze out (CFO) picture to the charm and bottom sector and thus advocating for a multi-CFO scenario in heavy ion collisions at LHC. The extracted $T_{F}$ was found to be roughly unchanged with the collision centrality with a tendency of a very mild drop towards most central collisions indicating a slight earlier freeze-out  in peripheral collisions or a weak contamination by spectator nucleons. The near constancy of $T_{F}$ across various centrality bins was considered as an evidence for thermalization, in a sense that initial state information is forgotten. 
	
	These results however suffered from the large uncertainty associated with the then available data, forbidding one to make any mature claim. Analysis of the then available preliminary charmonium data from CMS collaboration, in the same model framework, after removing non-prompt contribution from $b$-hadron decays gives $T_{f} = 159 \pm 31$ MeV at central rapidity and $265 \pm 59$ MeV at forward rapidity. The significant  difference in $T_f$ at the two rapidity intervals is attributed to the different $p_{T}$ acceptance of the corresponding data samples. Higher $p_{T}$ charmonia are likely to escape more easily from the fireball and thus having lesser probability of undergoing complete thermalization. Smaller $T_{F}$ for larger $p_{T}$ threshold thus indicates that charmonia which are sampled are yet to be equilibrated with the fireball. Decreasing $T_{F}$ towards more peripheral collisions, in both forward and central rapidity intervals lends support to this interpretation. The unavailability of suitable low-$p_{T}$ charmonia data samples prevented any robust determination of their freeze-out temperature. Compared to then the situation is much improved now in terms of availability of high precision data. Both CMS and ALICE collaborations have collected high statistics data for quarkonium production in p-p, p-Pb and Pb-Pb collisions. 
	
	This motivates us to perform a reinvestigation of quarkonium chemical freeze out analysis at LHC, by investigating their relative production yield in nuclear collisions. In this article, we have employed a semi-classical grand canonical approach and analyzed the relative yields of different bottomonia and charmonia states produced in Pb-Pb, p-Pb and high multiplicity p-p collisions at LHC. We have also systematically investigated the dependence of chemical freeze-out  temperature on the collision centrality. The article is organized in the following way. Section II gives a brief introduction to statistical model used in our calculation. The analysis of quarkonium yield ratios for different collision systems and at different energies within our model framework is performed in Section III. The main results of our investigation are summarized in section IV. 
	

	\section{Brief description of the model}
	
	The model used here is a thermal statistical model based on semiclassical grand canonical approach. The model is similar to the famous Hadron Resonance Gas (HRG) model that has been very successful in describing the particle abundances and thereby extracting the thermodynamic parameters of the fireball at chemical freeze-out~\cite{Braun-Munzinger:1994ewq, Cleymans:1999st, Becattini:2005xt, Andronic:2008gu, Vovchenko:2018fmh, Sarkar:2019oyo}. The partition function is constructed considering a grand canonical ensemble of non-interacting system of particles obeying quantum statistics. In this model, the primordial yield ($N^p$) of the $i^{th}$ hadron at zero chemical potential ($\mu =0$), can be obtained from the following equation~\cite{Andronic:2012ut, Sarkar:2017ijd},
	\begin{equation}
		\label{equ:n_idHRG}
		N^{p}_{i}=\frac{g_i V}{(2\pi)^3} \int d^3\textbf{p}\, \frac{1}{\exp\left(\frac{1}{T}\sqrt{m_i^2+\textbf{p}^2}\right)\pm 1}
	\end{equation}
	where, the sign, `$\pm$', corresponds to the fermion and boson respectively and other symbols have their usual meaning.
	
	The measured yield of $i^{th}$ hadron also contains the contribution of decay feed-down from heavier resonances. So, the total multiplicity of the $i^{th}$ hadron is given by the sum of the primordial ($N^p$) yield as well as the feed-down contribution from heavier resonance states, i.e.,
	\begin{equation}
		N^{t}_{i}=N^{p}_{i}+\sum_j N^p_j \textrm{ x } B.R_{j \rightarrow i}.
	\end{equation}
	where, $B.R_{j \rightarrow i}$  is the branching ratio of the respective decay channel. We have taken all the quarkonium resonance states and the different branching ratios as available in~\cite{ParticleDataGroup:2020ssz}. Note that we are concerned about the quarkonium states only and therefore all the hadronic states of mass below quarkonium do not contribute to the present calculation. 
	
	
	Another important point of consideration is that experiments do not cover the entire phase space and are rather restricted by different detector acceptance and other experimental constraints. These restrictions result in limited $p_T$ and $\eta$ coverage of the detectors. Hence, for a realistic comparison between the theoretical model estimation with experimental results, it is imperative to implement kinematic cuts in the model calculation. To incorporate experimental acceptance effects, we express Eq.~\eqref{equ:n_idHRG} as \cite{Garg:2013ata, Alba:2014eba},
	\begin{equation}\label{final_yield}
		N^{p}_{i}=\frac{g_i V}{4\pi^2} \!\int\! d\eta\,dp_T\, \frac{p_T^2 \cosh\eta }{\exp\!\left[\frac{1}{T}\sqrt{p_T^2\cosh^2\eta + m_i^2} \right]\pm 1} , 
	\end{equation}
	where the limits of the integration, $\eta_{\rm min}\leq\eta\leq\eta_{\rm max}$ and $p^{\rm min}_T\leq p_T \leq p^{\rm max}_T$, are determined from experimental kinematic cuts. Unless explicitly mentioned, these kinematic cuts are implemented in all of the analyses. 

	
	\section{Analysis Results}
	
	In this section, we separately study relative quarkonium yields in heavy ion collisions and in small systems (p-p and p-A collisions) at LHC.
	
	\subsection{A-A System}
	
	We begin with the thermal analysis of the relative quarkonium yields in Pb-Pb collisions at LHC. The data sets for various quarkonium states used in our analysis are summarized in Tab.~\eqref{Tab:PbPbSystem} and correspond to $J/\psi (1S)$ and $\psi (2S)$ states in the charmonia sector and $\Upsilon (1S)$, $\Upsilon (2S)$ and $\Upsilon(3S)$ states in the bottomonia sector, all reconstructed via their dimuon ($\mu^{+}\mu^{-}$) decay channel. In addition to the available LHC data, recently published STAR data on bottomonium production in $\sqrt{s_{NN}}=200$ GeV Au-Au collisions at RHIC and the available $\psi$-to-$J/\psi$ ratio in $\sqrt{s_{NN}}=17.3$ GeV Pb-Pb collisions measured by NA50 collaboration at SPS are also examined for completeness. For consistency check, it is imperative to compare our model analysis with the previously obtained results~\cite{Gupta:2014ova}. 
	
\begin{figure}[t]
		\centering
		\includegraphics[width=\linewidth]{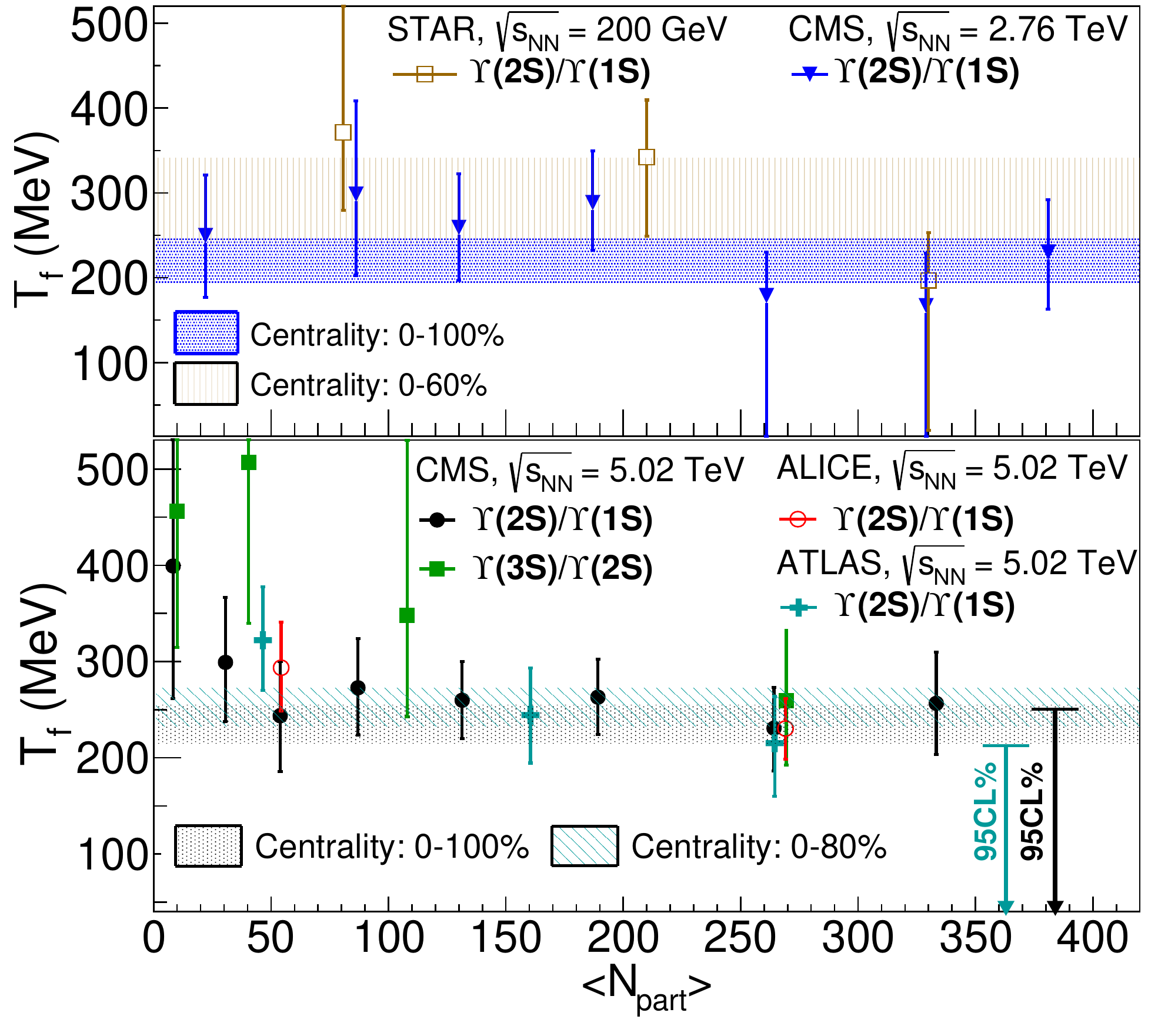}
		\caption{The centrality dependence of the freeze-out temperature ($T_f$) of the $\Upsilon$ mesons, in the A-A collision systems, extracted using different relative yields of the bottomonium states measured by different ultra-relativistic heavy ion  experiments. Upper panel: $T_f$ extracted from data collected by CMS collaboration at LHC, in $\sqrt{s_{NN}}$ = 2.76 TeV Pb-Pb collisions and STAR collaboration at RHIC in $\sqrt{s_{NN}}$ = 200 GeV Au-Au collisions. Lower panel: $T_f$ is extracted from the data collected by CMS, ALICE and ATLAS collaborations in $\sqrt{s_{NN}}$ = 5.02 TeV Pb-Pb collisions. The band for both panels represents $T_f$ for centrality integrated events. The references and kinematic acceptance of the corresponding experimental data are given in Tab.~\eqref{Tab:PbPbSystem}.}
		\label{fig:Upsilon_PbPb}
	\end{figure}

	We thus start reporting our results of $\varUpsilon$ production at mid-rapidity, measured by the CMS collaboration, in $\sqrt{s_{NN}}=2.76$ TeV Pb-Pb collisions~\cite{CMS:2012gvv,CMS:2013jsu}. As previously argued, $R_{AA}$ or double ratio are not useful quantities to perform a thermal analysis of the heavy mesons produced in these collisions. Instead, to extract thermal parameters, like freeze-out temperature employing the thermal model, one needs a single ratio $r_{PbPb}[h_2]=N_{PbPb}[h_2]/N_{PbPb}[h_1]$, i.e., relative yields of different hadronic states produced in the same collision system. We have used the Pb-Pb single ratios of different bottomonium states, $r_{PbPb}[\Upsilon(2S)]=N_{PbPb}[\Upsilon(2S)]/N_{PbPb}[\Upsilon(1S)]$, as reported in~\cite{CMS:2013jsu} to extract the centrality dependence of freeze-out temperature, $T_{f}$. The results of our analysis are listed in Tab.~\eqref{Tab:Analyes_Upsilon_PbPb} and displayed in the upper panel of Fig.\eqref{fig:Upsilon_PbPb}. Our estimated value of $T_f$, extracted from the centrality integrated relative yield comes out to be, $221_{-22}^{+21}$ MeV, in close agreement with the previous result. Also, as found earlier, no strong centrality dependence of $T_f$ is observed, apart from a mild drop towards more central collisions. The errors in the extracted $T_{f}$ values are estimated from calculations with the bottomonia yields varied between given errors. The most recent measurements of $\Upsilon$ production in $\sqrt{s_{NN}}=2.76$ TeV Pb-Pb collisions by CMS collaboration is reported in Ref.~\cite{CMS:2016rpc}, in the same kinematic domain but with improved statistics and precision compared to their previous run~\cite{CMS:2012gvv}. Inline with previous measurements, a strong centrality-dependent suppression is observed in Pb-Pb relative to p-p collisions, by factors of up to approximately $2$ and $8$, for the $\Upsilon(1S)$ and $\Upsilon(2S)$ states, respectively. The corresponding $T_f$ extracted from the centrality integrated yield ratios comes out to be $201^{+21}_{-23}$ MeV.  A centrality differential measurement is though not possible due to the difference in the published centrality bins for the $\Upsilon(1S)$ and $\Upsilon(2S)$ states.
	
	The measurement of $\varUpsilon$ production at $\sqrt{s_{NN}}=5.02$ TeV Pb-Pb collisions is now made available by CMS, ALICE and ATLAS collaborations at LHC. While $r_{PbPb}[\Upsilon(2S)]$ is directly available from ALICE~\cite{ALICE:2020wwx}, CMS and ATLAS do not provide the single ratio, rather they present their results in terms of double ratios~\cite{CMS:2017ycw, CMS:2022rna, ATLAS:2022xso}. Hence in order to construct the corresponding $r_{PbPb}[\Upsilon(2S)]$ from CMS and ATLAS data sets, we multiply the double ratio with the pp single ratio, $r_{pp}[\Upsilon(2S)]$, which is obtained by integrating the differential cross sections as a function of rapidity and $p_T$ given in~\cite{CMS:2018zza} and~\cite{ATLAS:2022xso} by CMS and ATLAS collaborations respectively. The corresponding $T_{f}$ values extracted for various collision centralities are displayed in the lower panel of Fig.~\eqref{fig:Upsilon_PbPb}. The uncertainty in the estimated $T_{f}$ reduces for $\sqrt{s_{NN}}$ = 5.02 TeV compared to 2.76 TeV and it is more prominent for more central collisions. No significant collision energy dependence in $T_{f}$ has been observed and the overall trend namely the centrality independence and modest decrease of freeze-out temperature for the most central collisions remains the same for all analyzed data sets as well. 
	
	We note that the centrality integrated $T_f$ value as obtained from ATLAS data set is slightly higher than that extracted from ALICE and CMS. However the integrated data samples available from ATLAS corresponds to $0 - 80 \%$ centrality whereas other two experiments published their results for $0 - 100 \%$ collision centrality. It is also interesting to take a note of the fact that the ALICE collaboration performed measurements at the forward rapidity region ($2.5 < y < 4.$), whereas the CMS ($|y| < 2.4$) and ATLAS ($|y| < 1.5$) experiments have collected data at mid-rapidity. However the corresponding $T_f$ values from ALICE data agree within uncertainty to that extracted from CMS and ATLAS mid-rapidity data, indicating the nearly rapidity independence of the freeze-out temperature. This is possibly a direct consequence of boost invariance of the produced fireball at such high collision energy. 
	
	Moreover, the CMS collaboration has also reported the double ratio of excited states, $\varUpsilon(3S)/\varUpsilon(2S)$, in 5.02 TeV Pb-Pb collisions. We have also utilized this piece of information to extract the corresponding $T_f$. The number of data points are smaller and associated error bars are larger due to reduced statistics. However it is conspicuous from the lower panel of Fig.~\eqref{fig:Upsilon_PbPb} that though $T_f$ extracted from yield ratios of different states are different at low multiplicity regions, they match within errors, at the high multiplicity region pointing towards a single freeze-out scenario of all bottomonium states in central Pb-Pb collisions. It is also very interesting to note that the application of the detector acceptance cut does not have any effect on the resulting $T_f$ values of the analyzed bottomonia states. All the collaborations have collected the $\varUpsilon$ data samples down to $p_{T}=0$, though the upper limit of the $p_T$ is different. This presumably means that the thermalization of these heavy resonances is dominated by the low momentum states, which spend more time inside the fireball. 
	
	For an explicit verification of this fact, we have set different upper limits of $p_T$ in our model (see Eq.~\eqref{final_yield}) and found that the upper limit of $p_T$ beyond 3.5 GeV does not influence the resulting $T_f$. We finish this discussion with the preliminary analysis of the results on $\Upsilon$ production in $\sqrt{s_{NN}}=200$ GeV AuAu collisions at mid rapidity, recently made available by STAR collaboration at RHIC~\cite{STAR:2022rpk}. Instead of  directly providing $r_{AuAu}[\Upsilon(2S)]$, data are published in terms of separate centrality dependence of $R_{AA}$ for $1S$ and $2S$ states. Hence to construct the desired single ratio, $R_{AA}[\Upsilon (2S)]$ is first divided by $R_{AA}[\Upsilon (1S)]$ to get the double ratio, and then the single ratio is obtained by multiplying the pp single ratio given in Ref.~\cite{Zha:2013uoa}. The results are included in the upper panel of Fig.\eqref{fig:Upsilon_PbPb}. Though for central collisions the corresponding $T_f$ values are in close agreement  at $\sqrt{s_{NN}}=200$ GeV and $\sqrt{s_{NN}}=2.76$ TeV, the centrality integrated value for $0 - 60 \%$ STAR data is larger than $0 - 100 \%$ ALICE data.  

	\begin{table}[ht]
		\centering
		
		\begin{tabular}[t]{lcccc}
			\hline
			& \multicolumn{4}{c}{Bottomonium}\\
			\hline
			System&$\sqrt{s_{NN}}$ &Expt.&Ref.& Kinematic acceptance\\
			\hline
			Au-Au&0.2&STAR&\cite{STAR:2022rpk}&$  p_T<10$, $ |y|<1.0$ \space  \space\space  \space\space   \\ 
			Pb-Pb&2.76&CMS&\cite{CMS:2012gvv,CMS:2013jsu}&$  p_T>0$, $ |y|<2.4$ \space  \space\space  \space\space   \\ 
			Pb-Pb&5.02&CMS&\cite{CMS:2017ycw, CMS:2018zza, CMS:2022rna}&$ p_T<30$, $ |y|<2.4$\space  \space \space  \space \\
			
			Pb-Pb&5.02&ALICE&\cite{ALICE:2020wwx}&$ p_T<15$, $ 2.5<y<4$\\
			\hline
			
			Pb-Pb&5.02&ATLAS&\cite{ATLAS:2022xso}&$ p_T<30$, $ |y|<1.5$\\
			\hline
			& \multicolumn{4}{c}{Charmonium}\\
			\hline
			Pb-Pb&0.017&NA50&\cite{NA50:2006yzz}&$p_T>0$, $0<|y|<1.$\\
			Pb-Pb&2.76&CMS&\cite{CMS:2014vjg, CMS:2012tva}& \vtop{\hbox{\strut $ 3<p_T<30$, $ 1.6<y<2.4$}\hbox{\strut $ 6.5 <p_T<15$, $|y|<1.6$}} \\
			Pb-Pb&2.76&ALICE&\cite{ALICE:2015jrl}&$p_T<15$, $2.5<|y|<4$\\	
			Pb-Pb&5.02&CMS&\cite{CMS:2016wgo, CMS:2017uuv}&$3<p_T<30$, $|y|<1.6$\\
			Pb-Pb&5.02&ALICE&\cite{CMS:2016wgo, CMS:2017uuv, ALICE:2022jeh}&$p_T<12$, $2.5<|y|<4.$\\
			
			\hline
		\end{tabular}
		\caption{Summary of the analyzed quarkonia data corpus for heavy-ion collisions, from different experiments at SPS, RHIC and LHC, along with their kinematic coverage. The freeze-out temperatures are extracted from the excited-to-ground state yield ratios. All the $\sqrt{s_{NN}}$ and $p_T$ are in TeV and GeV/c units, receptively.}
		\label{Tab:PbPbSystem}
	\end{table}
	\begin{figure*} 
		\centering\includegraphics[width=0.8\textwidth]{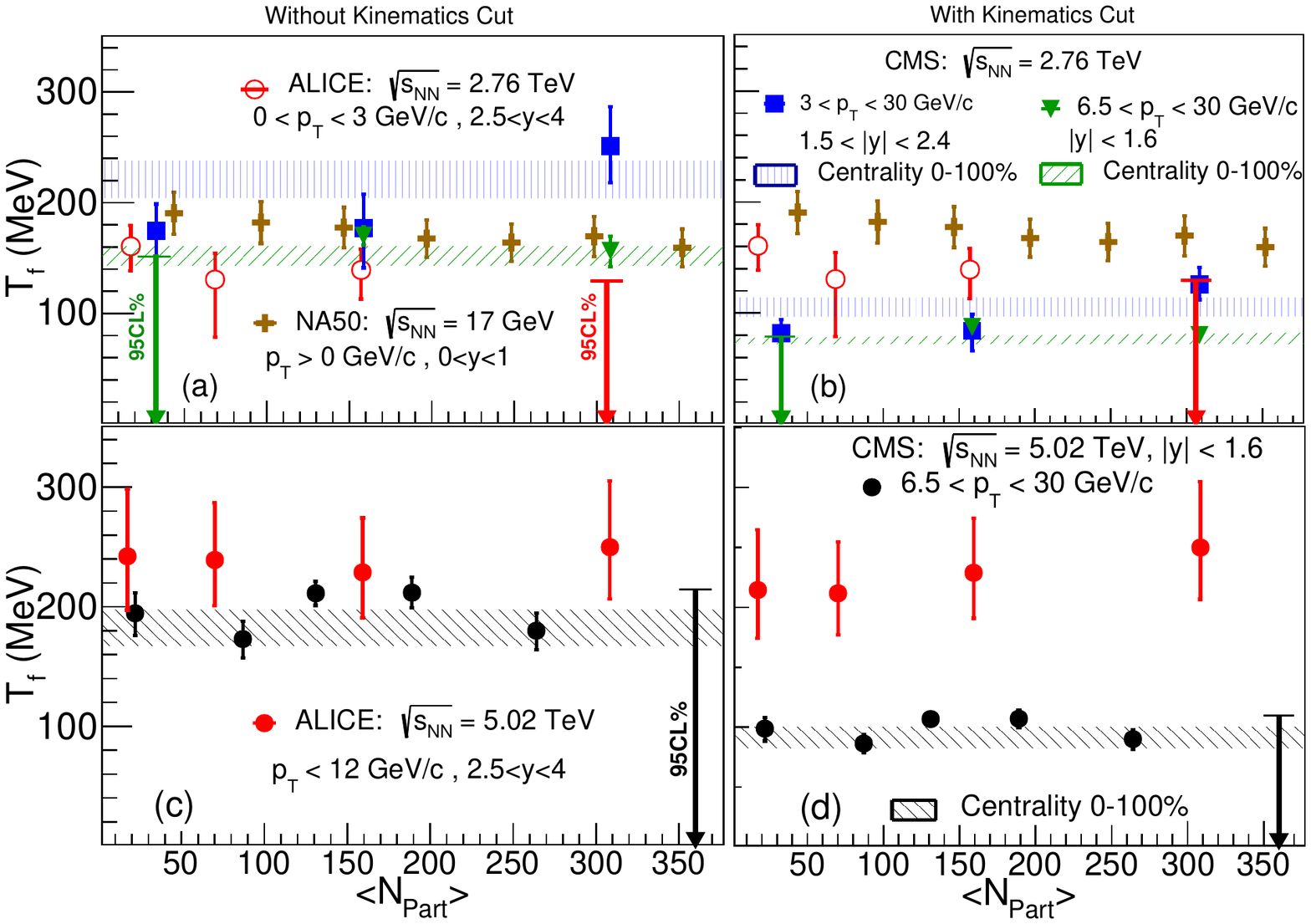}
		\caption{The centrality dependence of freeze-out temperature for charmonia states without (left) and with (right) acceptance correction effect, extracted using $r_{prompt}[\psi (2S)]$, in the Pb-Pb collisions at $\sqrt{s_{NN}}$ = 2.76 and 5.02 TeV Pb-Pb collisions measured by ALICE and CMS collaborations at LHC. For completeness, $T_f$ values estimated in $\sqrt{s_{NN}}$ =17.3 GeV Pb-Pb collisions, as measured by NA50 collaboration at SPS is also shown. The band represents $T_{F}$ for centrality integrated events. The red bands corresponding to ALICE 2.76 TeV and ALICE 5.02 TeV are absent due to to the unavailability of corresponding the data for centrality integrated yields from ALICE collaboration.}
		\label{fig:Jpsi_PbPb}
	\end{figure*}

	\begin{table}[ht]
		\centering
		\begin{tabular}[t]{clccc}
			\hline
			& \multicolumn{4}{c}{Bottomonium}\\
			\hline
			$\sqrt{s_{NN}}$&$<N_{part}>$&Collaboration& $T_f$(WOAC)&$T_f$(WAC)\\
			
			\hline
			\multicolumn{5}{c}{Kinematic Range: $ p_T<10$ GeV/c, $ |y|<1.0$ }\\
			\hline
			
			0.20 &88  &STAR& $371.35^{+117}_{-92}$& $371.35^{+117}_{-92}$\\ 
			0.20 &210  &STAR& $342.16^{+116}_{-93}$& $342.16^{+116}_{-93}$\\
			0.20  &335  &STAR& $196.75^{+86}_{-177}$& $196.75^{+86}_{-177}$\\

			0.20 &CI (0-60\%)     &STAR& $288.43^{+53}_{-49}$& $288.43^{+53}_{-49}$\\
			\hline
			\multicolumn{5}{c}{Kinematic Range: $ p_T>0$ GeV/c, $ |y|<2.4$ }\\
			\hline
			
			2.76 &22  &CMS& $249.70^{+71}_{-73}$& $249.70^{+71}_{-73}$\\ 
			2.76 &86  &CMS& $298.56^{+110}_{-96}$& $298.56^{+110}_{-96}$\\
			2.76  &130  &CMS& $259.34^{+63}_{62}$& $259.34^{+63}_{62}$\\
			2.76 &187  &CMS& $288.59^{+61}_{-56}$& $288.59^{+61}_{-56}$\\
			2.76 &261   &CMS& $178.96^{+51}_{-178}$& $178.96^{+51}_{-178}$ \\
			2.76 &329   &CMS& $167.12^{+62}_{-167}$& $167.12^{+62}_{-167}$\\
			2.76 &381   &CMS& $230.27^{+61}_{-67}$&$230.27^{+61}_{-67}$\\
			
			2.76 &CI (0-100\%)     &CMS& $221^{+26}_{-27}$& $221^{+26}_{-27}$\\
			\hline
			\multicolumn{5}{c}{Kinematic Range: $p_T<30$ GeV/c,  $ |y|<2.4$ }\\
			\hline
			5.02 &8.3    &CMS& $399.26^{+208}_{-138}$& $399.26^{+208}_{-138}$\\
			5.02 &30.6   &CMS& $299.13^{+68}_{-61}$& $299.13^{+68}_{-61}$\\
			5.02 &53.9   &CMS& $243.66^{+56}_{-57}$& $243.66^{+56}_{-57}$\\
			5.02 &87.0   &CMS& $272.72^{+51}_{-49}$& $272.72^{+51}_{-49}$ \\
			5.02 &131.4  &CMS& $259.76^{+40}_{-39}$& $259.76^{+40}_{-39}$\\
			5.02 &189.2  &CMS& $262.10^{+40}_{39}$& $262.10^{+40}_{39}$\\
			5.02 &264.2  &CMS& $230.73^{+42}_{-44}$& $230.73^{+42}_{-44}$\\
			5.02 &333.3  &CMS& $256.54^{+53}_{-53}$& $256.54^{+53}_{-54}$\\ 
			5.02 &384.3  &CMS& $<250.10$& $<250.10$  \\
			5.02 &CI (0-100\%)     &CMS& $235^{+19}_{-20}$& $235^{+19}_{-20}$\\
			\hline
			
			\multicolumn{5}{c}{Kinematic Range: $p_T<15$ GeV/c, $2.5<y<4$ }\\
			\hline
			5.02 &54.3&ALICE& $293.625^{+47}_{-44}$& $293.625^{+47}_{-44}$\\
			5.02 &269.1&ALICE& $230.435^{+31}_{-32}$&  $230.435^{+31}_{-32}$\\
			\hline
			
			\multicolumn{5}{c}{Kinematic Range: $p_T<30$ GeV/c, $|y|<1.5$ }\\
			\hline
			5.02 &46.5&ATLAS& $322.15^{+55}_{-52}$& $322.15^{+55}_{-52}$\\
			5.02 &160.496&ATLAS& $244.44^{+49}_{-50}$&  $244.44^{+49}_{-50}$\\
			5.02 &264.62&ATLAS& $215.12^{+47}_{-55}$& $215.12^{+48}_{-55}$\\
			5.02 &362.98&ATLAS& $<212.75 $&  $<212.76$\\
			5.02 &CI (0-80\%)&ATLAS& $253.53^{+45}_{-45}$&  $253.53^{+46}_{-45}$\\
			\hline
		\end{tabular}
		
		\caption{The freeze-out temperature of the bottomonium states produced in $\sqrt{s_{NN}}$ =200 GeV Au-Au collisions as measured by STAR collaboration at RHIC and in $\sqrt{s_{NN}}$ =2.76 TeV and 5.02 TeV Pb-Pb collisions at LHC as measured by ALICE, CMS and ATLAS collaborations at LHC. All the temperature and $\sqrt{s_{NN}}$ are in MeV and TeV units respectively. `WOAC' and `WAC' represent without and with acceptance correction, respectively. }
		\label{Tab:Analyes_Upsilon_PbPb}
	\end{table}

	\begin{table}[ht]
		\centering
		\begin{tabular}[t]{clccc}
			\hline
			& \multicolumn{4}{c}{Charmonium}\\
			\hline
			$\sqrt{s_{NN}}$&$<N_{part}>$&Collaboration& $T_f$(WOAC)&$T_f$(WAC)\\
			\hline
			
			\multicolumn{5}{c}{Kinematic Range: $p_T>0$ GeV/c, $0<|y|<1$ }\\
			\hline
			0.017 &43.72&NA50&$190.48^{+19}_{-19}$& $190.48^{+19}_{-19}$\\
			0.017 &96.18&NA50&$182.15^{+19}_{-19}$& $182.15^{+19}_{-19}$\\
			0.017 &146.65&NA50&$177.50^{+18}_{-18}$& $177.50^{+18}_{-18}$\\
			0.017 &196.86&NA50& $167.60^{+17}_{-17}$& $167.61^{+17}_{-17}$\\
			0.017 &248.33&NA50& $163.98^{+16}_{-17}$& $163.98^{+16}_{-17}$\\
			0.017 &298.54&NA50& $169.67^{+18}_{-18}$& $169.68^{+18}_{-18}$\\
			0.017 &351.89&NA50& $159.37^{+17}_{-17}$& $159.37^{+17}_{-17}$\\
			\hline

			\multicolumn{5}{c}{Kinematic Range: $3<p_T<30$ GeV/c, $1.6<|y|<2.4$ }\\
			\hline
			2.76 &32.8&CMS&$174.40^{+24}_{-27}$& $82.06^{+13}_{-14}$\\
			2.76 &158.6&CMS&$177.20^{+31}_{-36}$& $83.95^{+15}_{-18}$\\
			2.76 &308.4&CMS&$251.27^{+35}_{-33}$& $126.05^{+15}_{-14}$\\

			2.76 &CI&CMS& $220.10^{+17}_{-17}$& $105.41^{+9}_{-9}$\\
			\hline
			\multicolumn{5}{c}{Kinematic Range: $6.5<p_T<30$ GeV/c, $|y|<1.6$ }\\
			\hline
			2.76 &32.8&CMS& $<153.33$ &  $<78.72$\\
			2.76 &158.6&CMS& $170.11^{+8}_{-8}$& $87.41^{+5}_{-5}$\\
			2.76 &308.4&CMS& $156.55^{+13}_{-15}$& $80.09^{+6}_{-7}$\\

			2.76 &CI&CMS   & $151.69^{+9}_{-8}$& $77.75^{+6}_{-5}$\\
			\hline
			\multicolumn{5}{c}{Kinematic Range: $p_T<15$ GeV/c, $2.5<y<4$ }\\
			\hline
			2.76 &17.51&ALICE& $160.59^{+19}_{-22}$& $160.60^{+19}_{-22}$\\
			2.76 &68.6&ALICE&$130.57^{+24}_{-52}$&$130.58^{+24}_{-52}$\\
			2.76 &157.2&ALICE& $139.17^{+19}_{-26}$& $139.18^{+19}_{-26}$\\
			2.76 &308.1&ALICE& $<128.20$& $<128.32$\\
			\hline
			
			\multicolumn{5}{c}{Kinematic Range: $6.5<p_T<30$ GeV/c, $|y|<1.6$ }\\
			\hline
			5.02 &22&CMS& $194.54^{+17}_{-19}$& $98.78^{+9}_{-10}$\\
			5.02 &87&CMS& $172.98^{+15}_{-16}$& $86.35^{+7}_{-8}$\\
			5.02 &131&CMS& $211.31^{+10}_{-10}$&$ 106.78^{+6}_{-6}$\\
			5.02 &189&CMS& $211.83^{+13}_{-13}$&$ 107.05^{+7}_{-7}$\\
			5.02 &264&CMS& $179.98^{+15}_{-16}$&$ 90.05^{+8}_{-8}$\\
			5.02 &359&CMS& $<214.43$  & $<109.57$ \\
			5.02 &CI&CMS& $182.72^{+15}_{-16}$ & $91.50^{+9}_{-9}$ \\
			\hline
			
			\multicolumn{5}{c}{Kinematic Range: $ p_T<12$ GeV/c, $2.5<y<4$ }\\
			\hline
			5.02 &17&ALICE& $242.25^{+56}_{-45}$& $214.46^{+50}_{-40}$\\
			5.02 &70&ALICE& $239.14^{+48}_{-38}$& $211.38^{+43}_{-34}$\\
			5.02 &159&ALICE& $228.81^{+46}_{-38}$&$ 228.81^{+45}_{-38}$\\
			5.02 &308&ALICE& $249.86^{+55}_{-43}$&$ 249.86^{+55}_{-43}$\\
			\hline
		\end{tabular}
		
		\caption{Same as the caption of the Tab.~\eqref{Tab:Analyes_Upsilon_PbPb} but for charmonium  states.}
		\label{Tab:Analyes_Jpsi_PbPb}
	\end{table}

	Let us now turn to the available data in charmonium sector at LHC. $R_{PbPb}(J/\psi)$ and its systematics have been reported by the ALICE, ATLAS and CMS collaborations. However the comparison of $\psi(2s)$ and $J/\psi$ is yet to be available from ATLAS. For the present study, we have thus analyzed the data from both CMS and ALICE collaborations at $\sqrt{s_{NN}} = 2.76$ TeV~\cite{ALICE:2015jrl,CMS:2014vjg, CMS:2012tva} and $\sqrt{s_{NN}}=5.02$ TeV~\cite{CMS:2016wgo, ALICE:2022jeh}. The relative yield of $\psi'$ with respect to $J/\psi$ is used to estimate the corresponding $T_f$. But such extraction should be performed on the prompt charmonium states free from the feed down contribution from $b$ quark decays.  The CMS collaboration, thanks to their excellent vertex resolution capabilities, has published the prompt yield of $J/\psi$ and $\psi'$ mesons in Pb-Pb collisions. Alike bottomonium measurements, CMS primarily provides double ratio in both 2.76 TeV~\cite{CMS:2014vjg} and 5.02 TeV~\cite{CMS:2017uuv} Pb-Pb collisions, except for the centrality integrated single ratio at 2.76 TeV, reported in~\cite{CMS:2012tva}. Hence, we need to use pp $p_{T}$-differential cross-section for prompt $\psi (2S)$~\cite{CMS:2018gbb} and $J/\psi$~\cite{CMS:2017uuv} mesons to calculate $p_{T}$-integrated pp single ratio. Subsequently we calculate prompt Pb-Pb single-ratio, $r_{PbPb}[\psi (2S)]= N_{PbPb}[\psi (2S)]/N_{PbPb}[J/\psi]$.
	The ALICE collaboration, on the other hand, does not report prompt yields of different charmonium states, but rather provides inclusive yield ratios. 
	
	For exclusion of the non-prompt contribution and construct $r_{prompt}[\psi(2s)]$ from the ALICE data, we follow the same prescription given in Ref.~\cite{Gupta:2014ova}
	\begin{equation}
		r_{prompt}[\psi (2S)]=r[\psi (2S)]\Big(\frac{1-af_B}{1-f_B}\Big).
	\end{equation}
	where `$f_B$' and `$af_B$' are respectively the fraction of $J/\psi$ and $\psi(2S)$ coming from weak decay of $B$-hadrons. $f_B$ is experimentally measured in p-p collisions. The anticipated value of $a$ is based on the ratio of partial widths of the decays $B^{+} \rightarrow \psi(2S)\pi$ and $B^{+} \rightarrow J/\psi\pi$. In our calculation, we take a central value $\sim$ 0.5 and vary it from 0 to 1 that contributes to the uncertainty in estimated prompt ratio. At $\sqrt{s_{NN}}$ = 2.76 TeV, in the forward rapidity regime, $f_B$ is found to be $10.7\pm 4.8 \pm 2.5 \%$, in the kinematic range $1.5<p_T<4.5$~\cite{ALICE:2015nvt}. ALICE recently measured the non-prompt contribution of $J/\psi$ down to $p_T$ = 1 GeV at $\sqrt{s_{NN}}$ = 5.02 TeV \cite{ALICE:2022fb}. It appeared that $f_B$ has a strong $p_T$ dependence. In the $p_T$ range of our interest ($p_T < 12$ GeV), $f_B$ varies from 10 to 30\% \cite{ALICE:2022fb}. We take the mean value, i.e., 20\% in our calculation. The resulting variation in the $T_f$ is $\sim$ 7.5 \% due to the variation in  $f_B$ from 10 to 30\%, which is added to the uncertainty of the prompt ratio in our analysis. The centrality~\cite{ALICE:2015nvt} and rapidity~\cite{CMS:2017uuv} independency of the $f_B$ has been considered throughout the analyses. For completeness, we have also analyzed the data available from NA50 collaboration at SPS on $\psi'$-to-$J/\psi$ ratio in $\sqrt{s_{NN}} = 17.36$ GeV Pb-Pb collisions~\cite{NA50:2006yzz}. However the weak decay contribution is ignored due to negligible production of beauty hadrons at SPS.

	The analysis results are displayed in Fig.~\eqref{fig:Jpsi_PbPb} and tabulated in Tab.~\eqref{Tab:Analyes_Jpsi_PbPb}. For a quantitative comparison, we have calculated $T_f$ both, with and without experimental acceptance cuts. One may immediately note the difference between the $T_f$ values extracted from the CMS and ALICE data sets. 
	
	For CMS data corpus, which have a finite low-$p_T$ cut off, we observe that the acceptance correction significantly reduces the corresponding $T_f$ values at both collision energies. CMS measured charmonium production in $\sqrt{s_{NN}}$ = 2.76 TeV at forward ($ 1.5<|y|<2.4$) and mid-rapidity ($|y|<1.6$) regions with $p_T$ coverage of $3 < p_{T} < 30$ GeV/$c$ and $6.5 <p_{T} <30$ GeV/$c$, respectively and in $\sqrt{s_{NN}}=5.02$ TeV only at mid-rapidity region where the $p_T$ coverage is same as that of previous one. For mid-rapidity (larger $p_T$ threshold) the corresponding $T_f$ values are smaller than that at forward rapidity. For a given low $p_T$ threshold, $T_f$ increases for more central collisions. These observations are inline with previous findings where it was attributed to different $p_T$ acceptances. The selected high $p_{T}$ mesons would spend lesser time inside the medium and thus less likely to be thermalized with the fireball. Note that in the previous work the acceptance effect was not taken into account. Explicit incorporation of the finite acceptance correction in our calculations and resulting reduction in the freeze-out temperature directly manifests the effect of the large minimum $p_T$ threshold of the analyzed charmonia data sets. A thermal interpretation of the charmonium data sampled by CMS collaboration is thus possibly incorrect. One may take note of the fact that the recent measurements by CMS collaboration, of $J/\psi$ production inside jets in $\sqrt{s}=8$ TeV p-p collisions led them to conclude jet fragmentation as the dominant source of high $p_{T}$ ($E_{J/\psi} > 15$ GeV) prompt $J/\psi$ at mid rapidity~\cite{CMS:2019ebt}. 
	
	\begin{figure}
		\includegraphics[width=0.45\textwidth]{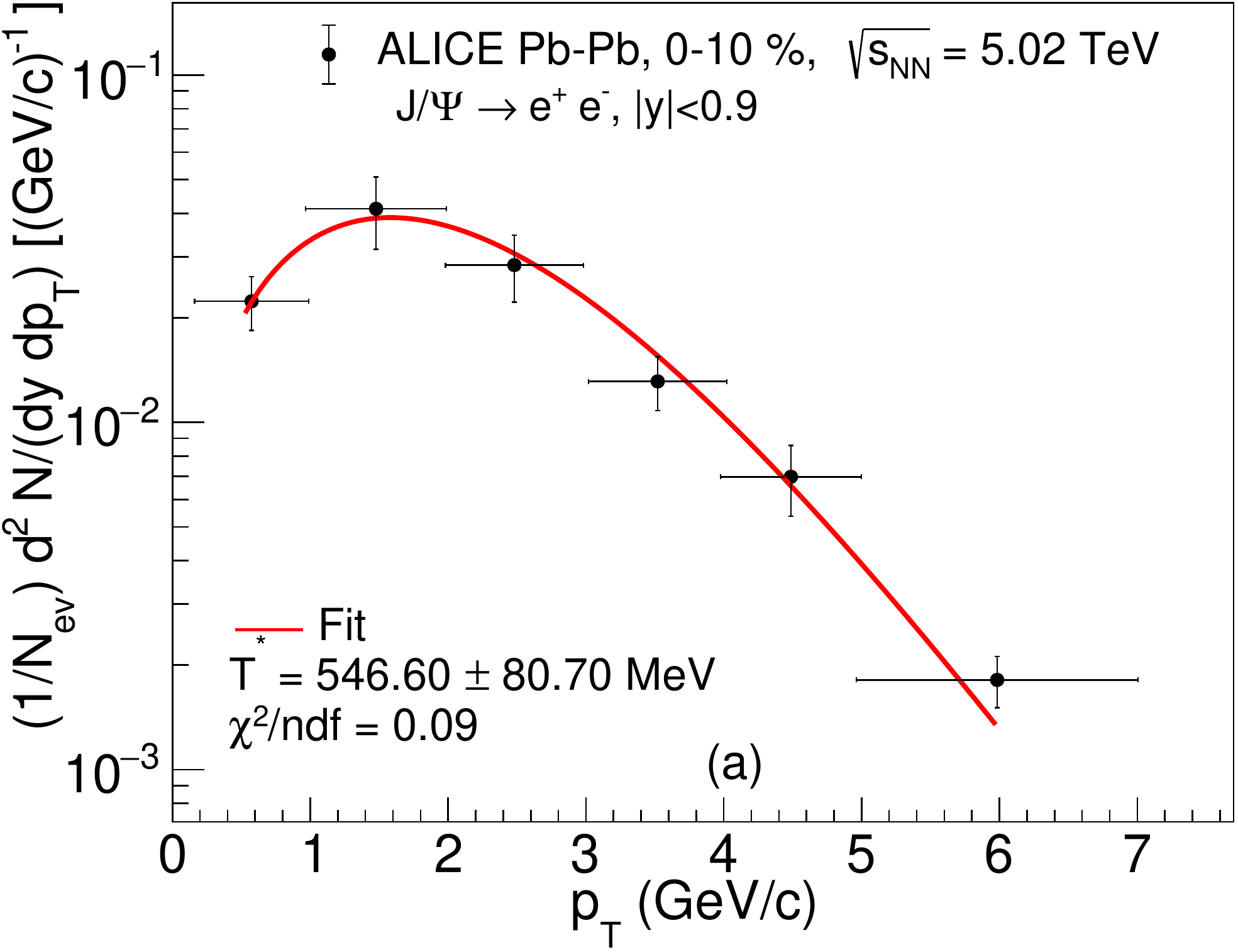}
		\includegraphics[width=0.45\textwidth]{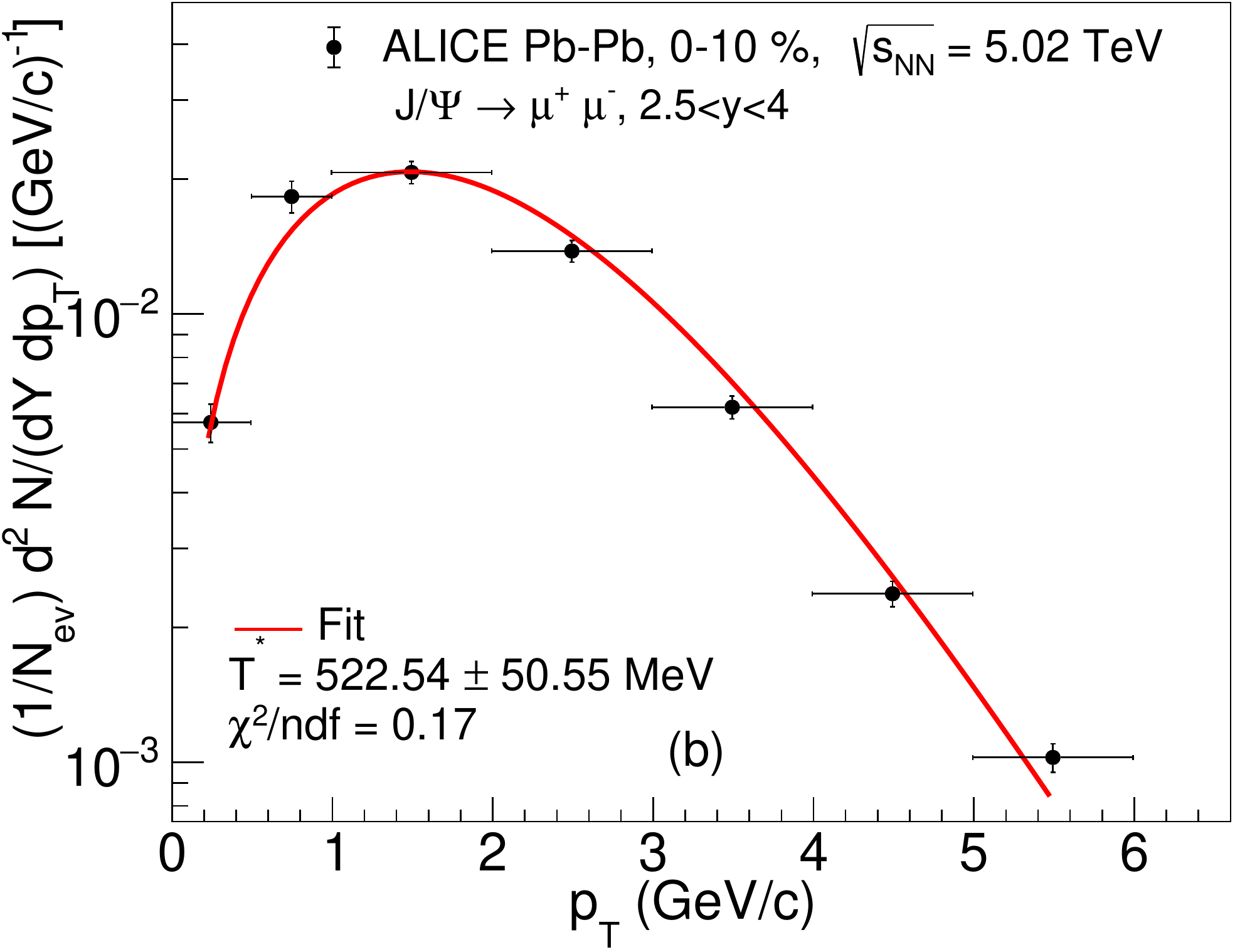}
		\vspace{-0.2cm}
		\caption{Thermal model fits to the transverse momentum ($p_{T}$) spectra of $J/\psi$ mesons in $\sqrt{s_{NN}}=5.02$ TeV most central ($ 0 - 10 \%$) Pb-Pb collisions at (a) mid-rapidity ($|y| < 0.9$) \cite{ALICE:2023gco}, and (b) forward rapidity ($2.5 < y < 4.0$) \cite{ALICE:2019lga}. The fit are restricted upto $p_{T} \sim 6$ GeV/$c$. For both the data sets the fit qualities are similar (comparable values of $\chi^{2}$ per degree of freedom ($\chi^{2}$/ndf) and the corresponding effective temperature ($T^{*}$) values match within errors.} 
		\label{pT_Jpsi}
	\end{figure}

	This claim gets further support from the analysis of now available ALICE data. For the ALICE data sample, measured at forward rapidity ($2.5 < y < 4.0$), the minimum threshold value of $p_T$ is as low as zero and the effect of acceptance correction on the resulting $T_f$ is absent. These data points thus give a genuine $T_f$ of the charmonium system at LHC. For most central collisions, the $T_{f}$ for charmonium states in 2.76 TeV Pb-Pb collisions, comes out to be around 140 MeV. Accounting for the related uncertainty, the value is in agreement with the freeze-out temperature of light hadrons. At 5.02 TeV the corresponding value of $T_f$ is around 250 MeV. Contrary to the CMS  results, a mild drop in $T_f$ observed with increasing collision centrality. The large recombination effect in the charmonium sector~\cite{Braun-Munzinger:2000csl, Thews:2000rj} may explain the high value of $T_f$ extracted from ALICE 5.02 TeV data~\cite{ALICE:2022jeh}. As has been shown in Ref.~\cite{ALICE:2022jeh}, the centrality dependence ratio is well explained by a model such as TAMU~\cite{Du:2015wha} that takes recombination into account during the plasma phase.  The effect of recombination is found to be stronger for $\psi(2S)$ than $J/\psi$, leading to larger values of the  yield ratio and hence the corresponding $T_f$. For two most peripheral bins of ALICE data at 5.02 TeV, we observe a sensitivity in the $T_f$ values on the detector acceptance. A low $p_T$ cut off of 300 MeV was set in the analyzed data for these two peripheral bins to get rid of charm contribution from photo production~\cite{ALICE:2022jeh}, which then affects the resulting $T_f$ values. 
	
	Before we move on, it will also be interesting to investigate the available data on transverse momentum ($p_{T}$) spectra of $J/\psi$ mesons in Pb-Pb collisions at LHC, following the thermal model description of Eq.~\ref{final_yield}. This would be an additional validation of our thermal model description with finite phase space coverage. However one needs to keep in mind that the particle multiplicities or their ratios probe the chemical freeze-out temperature ($T_{f}$), while the $p_T$ spectra probe the kinetic freeze-out temperature ($T_{kin}$). Also the effect of medium expansion is folded in the inverse slope parameter of the $p_{T}$ spectra and hence boosted thermal models like blast wave models are more suitable to analyze thermal distributions. Nevertheless we have tried to fit the $p_T$ distribution of the $J/\psi$ mesons in 5.02 TeV most central PbPb collisions at LHC, measured by ALICE collaboration, both at mid-rapidity ($|y| < 0.9$) and at forward rapidity ($2.5 < y < 4.0$), using Eq.~\ref{final_yield}. We restrict our fit up to $p_{T} ~ 6$ GeV/$c$ in order to remove other effects like ``corona" contribution and jet fragmentation \cite{Andronic:2019wva}. The fit results are displayed in Fig.~\ref{pT_Jpsi}. We obtain the so called ``effective'' temperatures $T^{*} = 546.60 \pm 80.70$ MeV at mid rapidity and $T^{*} = 522.54 \pm 50.55$ MeV at forward rapidity. The extracted $T^{*}$ values match within errors. The corresponding kinetic freeze-out temperatures can be obtained from the approximate relation $T^{*} \simeq T_{kin} + 1/2 m_{J/\psi}\beta^{2}$ \cite{Gorenstein:2001ti}, where $\beta$ denotes the average transverse velocity of the fireball. For heavy mesons like $J/\psi$, we assume $T_{kin} \simeq T_{f} \simeq 250$ MeV, as obtained in Tab.~\eqref{Tab:Analyes_Jpsi_PbPb}. We get mean transverse velocity $\beta=0.44$ at mid rapidity and $\beta = 0.42$ at forward rapidity, which is a reasonable estimate.

	
	\subsection{Small Systems (p-p, p-Pb)}

		\begin{table}[b]
		\centering
		\begin{tabular}[t]{lcccc}
			\hline
			& \multicolumn{4}{c}{Small System: p-p, p-Pb}\\
			\hline
			& \multicolumn{4}{c}{Bottomonium}\\
			\hline
			System&$\sqrt{s_{NN}}$ &Expt.&Ref.&Kinematic acceptance \\
			\hline
			p-p&2.76&CMS&\cite{CMS:2013jsu}&$p_T>0$ GeV/c, $|y|<1.93$\\
			p-p&7&CMS&\cite{CMS:2020fae}&$p_T>0$ GeV/c, $|y|<1.93$\\
			p-p&13&ALICE&\cite{Chowdhury:2019vzm}&$p_T>0$ GeV/c, $2.5<y<4$\\
			p-Pb&5.02&CMS&\cite{CMS:2013jsu}&$p_T>0$ GeV/c, $|y|<1.93$\\
			p-Pb&8.16&ALICE&\cite{ALICE:2019qie}&$p_T<15$ GeV/c \\
			\hline
			& \multicolumn{4}{c}{Charmonium}\\
			\hline

			p-p&5.02&ALICE&\cite{ALICE:2021qlw}& $0<p_T<12$\\  
			
			p-p&7&ALICE&\cite{ALICE:2014uja}& $0<p_T<12$\\
			p-p&8&ALICE&\cite{ALICE:2015pgg}& $0<p_T<12$\\
			p-p&13&ALICE&\cite{ALICE:2017leg}& $0<p_T<16$\\
			p-Pb&5.02&ALICE&\cite{ALICE:2016sdt}&\vtop{\hbox{\strut $ -4.46<y<-2.96$, $ 2.03<y<3.53$}\hbox{\strut .\hspace{18 mm}  $p_T<20$ GeV/c}}\\
			
			p-Pb&5.02&ALICE&\cite{ALICE:2014cgk}&\vtop{\hbox{$p_T<8$ GeV/c}}\\
			
			p-Pb&8.16&ALICE&\cite{ALICE:2020tsj}&\vtop{\hbox{\strut $ -4.46<y<-2.96$, $ 2.03<y<3.53$}\hbox{\strut .\hspace{18 mm}  $p_T<20$ GeV/c}}\\
			
			p-Pb&8.16&ALICE&\cite{ALICE:2020vjy}&$p_T<20$ GeV/c\\
			\hline
		\end{tabular}
		\caption{Same as the caption of Tab.~\eqref{Tab:PbPbSystem}, but for the p-p and p-Pb system.}
		\label{Tab:SSystem}
	\end{table}

		\begin{figure}[t]
		\centering
		\includegraphics[height=3.15in,width=3.3in]{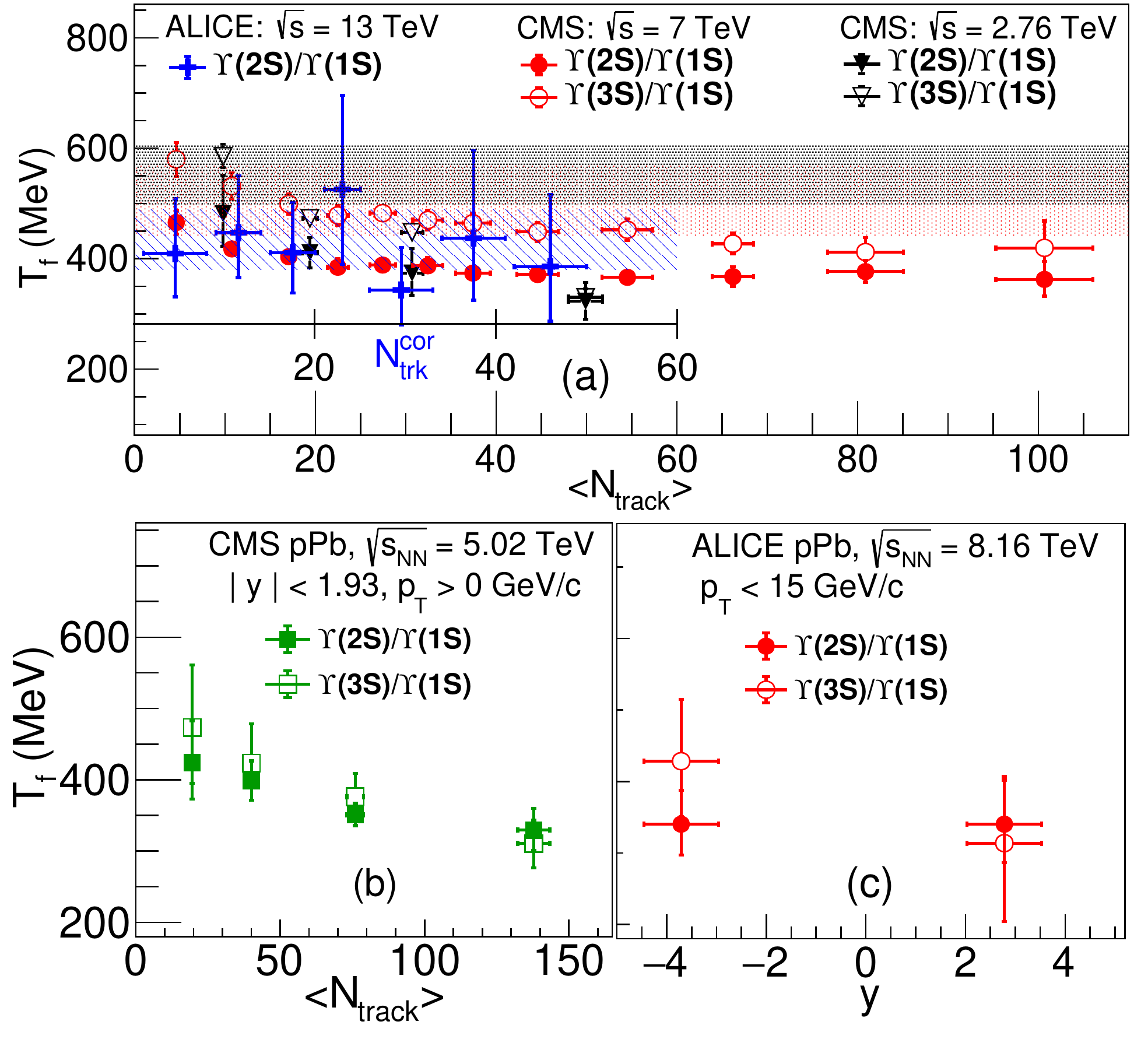}
		\caption{(a) (upper panel): The track multiplicity dependence of the bottomonia freeze-out temperature for the p-p collision systems at $\sqrt{s_{NN}}$= 2.76, 7 and 13 TeV. The band represents $T_f$ for minimum bias events. (b) (lower left panel): Same as the upper panel but for the p-Pb collisions at $\sqrt{s_{NN}}$= 5.02 TeV. (c) (lower right panel): Bottomonia freeze-out temperature as a function of rapidity for p-Pb collisions at $\sqrt{s_{NN}}$= 8.16 TeV. The freeze-out temperatures are extracted using the relative yields of the different bottomonium states. The references and kinematic acceptance of the corresponding experimental data are given in the Tab.~\eqref{Tab:SSystem} (see text for details).}
		\label{Fig:Upsilon_SSystem}
	\end{figure}
	
		Recently, some of the collective features of heavy ion collision have also been observed in the system formed in pp and p-Pb collisions contradicting the traditional wisdom. Some experimental findings that triggered the debate on the origin of the collectivity in small systems are long-range, near-side azimuthal correlation \cite{CMS:2010ifv, CMS:2015fgy, CMS:2012qk, LHCb:2015coe}, mass-ordering of elliptical flow \cite{CMS:2016fnw}, and $p_T$-dependent $v_2$ \cite{ATLAS:2015hzw} among the others. In view of this situation, it is interesting to test the thermal behavior of the quarkonium states produced in the p-p and p-Pb collisions with that of heavy ion collisions.
	
	The analyzed data corpus as available from CMS and ALICE collaborations are summarized in Tab.~\eqref{Tab:SSystem}. We begin with the analysis of the relative yields of bottomonia measured at different collision energies and for the p-p and p-Pb collision systems. For both collision systems, we have extracted $T_{f}$ for different track multiplicity classes (mimicking collision centrality) using both the ratios $r_{pp(Pb)}(\varUpsilon(2S)$ and $r_{pp(Pb)}(\varUpsilon(3S)$. The results are shown in Fig.\eqref{Fig:Upsilon_SSystem}. The upper panel shows the multiplicity-dependent freeze-out temperatures in the p-p collisions, extracted using the CMS data samples collected at $\sqrt{s_{NN}}= 2.76$ TeV \cite{CMS:2013jsu}, $\sqrt{s_{NN}}=7$ TeV \cite{CMS:2020fae} and also ALICE data at $\sqrt{s_{NN}}=13 $ TeV \cite{Chowdhury:2019vzm}. Note that CMS presents their results in terms of average $N_{track}$, whereas ALICE prefers to use corrected tracklets, $N_{trk}^{cor}$. The $T_f$ extracted from both types of relative yields has not shown any significant $\sqrt{s}$ dependence, though there is a tendency of decreasing $T_{f}$, extracted from minimum bias events, with increasing $\sqrt{s}$. 
	
		\begin{figure} 
		\centering
		\includegraphics[width=\linewidth]{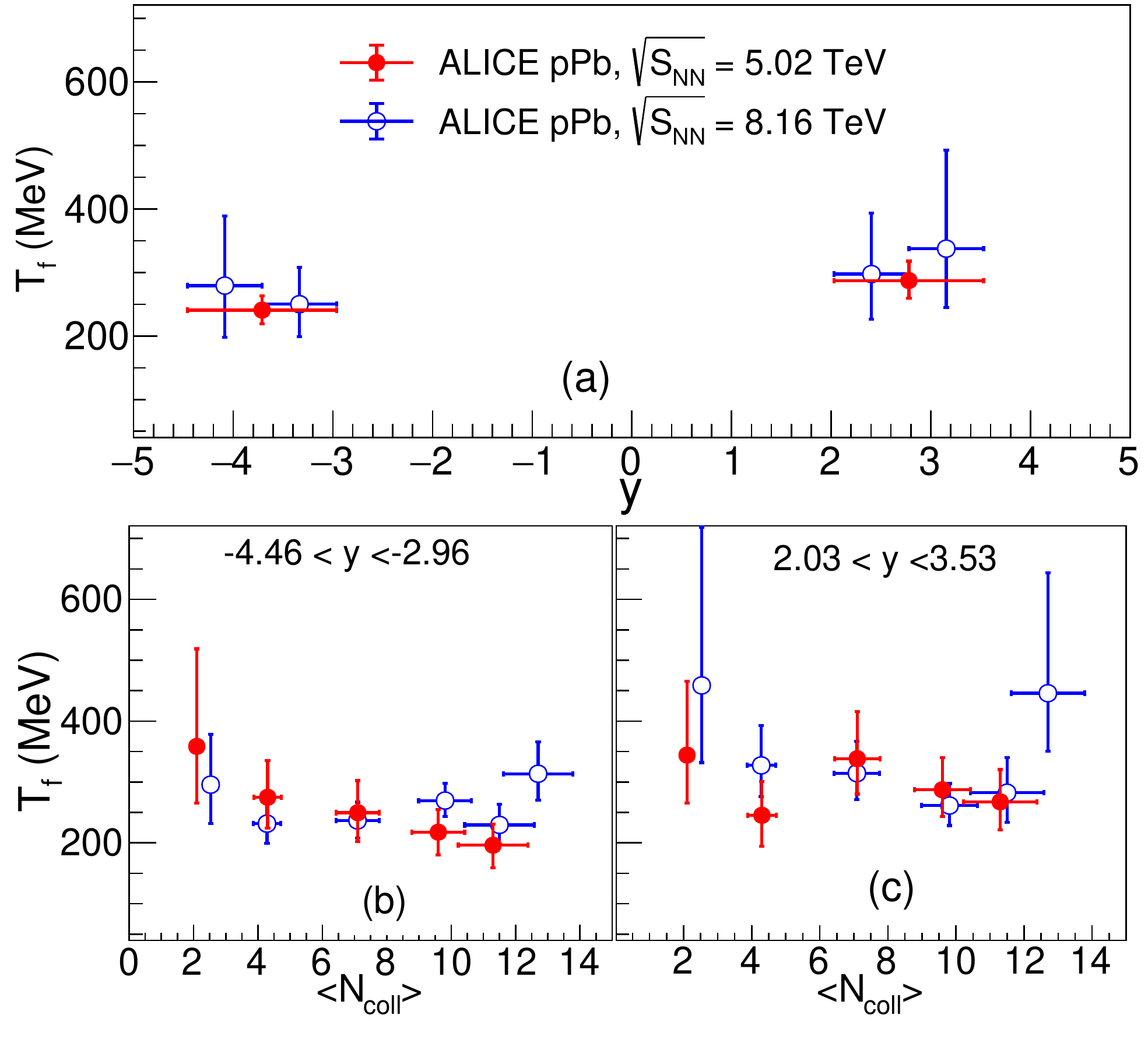}
		\caption{(a) (upper panel): Rapidity dependence of the charmonia freeze-out temperature in p-Pb collisions at $\sqrt{s_{NN}}$= 5.02 and 8.16 TeV. (b) (lower left panel): Centrality dependence of the charmonia freeze-out temperature in  $\sqrt{s_{NN}}$= 5.02 and 8.16 TeV p-Pb collisions at backward hemisphere. The centrality is estimated in terms of average number of binary collisions ($<N_{coll}>$). (c) (lower right panel): same as (b) but at forward hemisphere. The freeze-out temperatures are extracted using the relative yields of the different charmonium states. The references and kinematic acceptance of the corresponding experimental data are given in the Tab.~\eqref{Tab:SSystem} (see text for details).}
		\label{fig:Jpsi_pPb}
		
	\end{figure}
	
	\begin{figure}[b!]
		\centering
		\includegraphics[width=\linewidth]{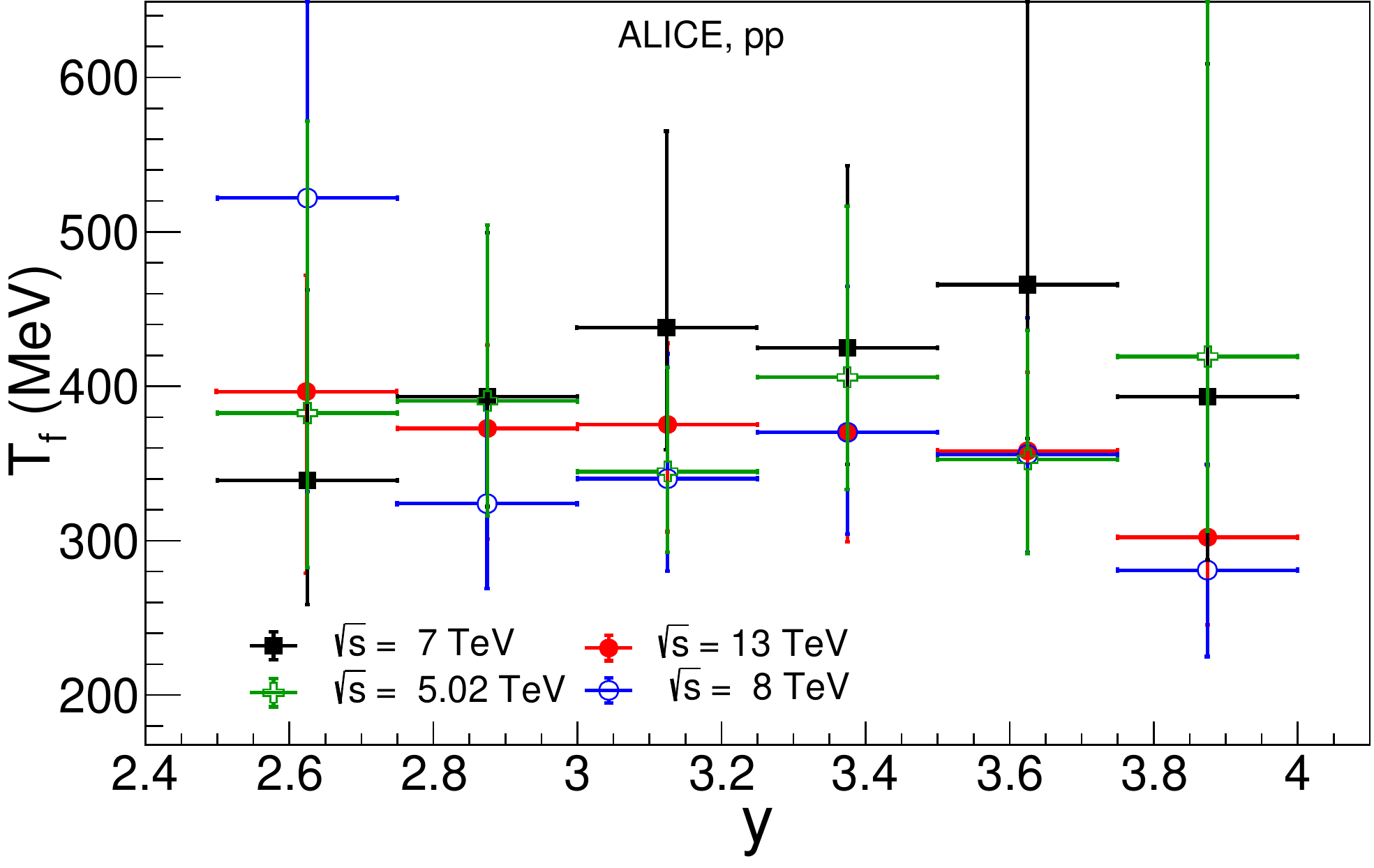}
		\caption{Rapidity dependence of the charmonium freeze-out temperatures in the forward regions, in p-p collisions at different center-of-mass energies ($\sqrt{s}$). The freeze-out temperatures are extracted using the relative yields of the different charmonium states. The references and kinematic acceptance of the corresponding experimental data are given in the Tab.~\eqref{Tab:SSystem} (see text for details).}
		\label{fig:Jpsi_pp}
		
	\end{figure}
        
\begin{table*}[bt]
		\begin{center}
		\label{Tab:single_fit}
		\begin{tabular}{|c|c|c|c|c|c|}
			\hline
			\multicolumn{6}{|c|}{ Thermal Fit Results}\\
			\hline
			System & Collision Energy &  Yield Ratio   & $\chi^2$/ndf   &    \thead {$T_{f}$ (Simultaneous fit of different  \\  centrality bin data) (MeV)} &\thead { $T_{f}$  (centrality integrated \\  yield ratio) (MeV)} \\
			
			\hline
			
			Pb-Pb
			&2.76 TeV &$\Upsilon(2S)/\Upsilon (1S)$ &0.82   & 229.71 $\pm$ 22.98 &  221.0 $\pm$ 26.0    \\[+1mm]
			\hline
			
			Pb-Pb
			&5.02 TeV &$\Upsilon(2S)/\Upsilon (1S)$ &0.34   & 258.31 $\pm$ 17.54 &  235.0 $\pm$ 19.0   \\[+1mm]
			\hline

			p-p
			&2.76 TeV &$\Upsilon(2S)/\Upsilon (1S)$ &2.76   & 390.45 $\pm$ 18.38 &  --   \\[+1mm]
			\hline

			p-p
			&7 TeV &$\Upsilon(2S)/\Upsilon (1S)$ &1.66   & 387.13 $\pm$ 14.28 &  500 $\pm$ 59   \\[+1mm]
			\hline

			\hline
		\end{tabular}%
                \end{center}
                \caption{Results of single parameter ($T_{f}$) thermal model fit to the excited ($\Upsilon (2S)$) to ground state ($\Upsilon(1S)$) ratio of the bottomonia for various centrality bins in $\sqrt{s_{NN}} = 2.76$ TeV and $\sqrt{s_{NN}} = 5.02$ TeV Pb-Pb and in $\sqrt{s_{NN}} = 2.76$ TeV and $\sqrt{s_{NN}} = 7$ TeV pp collisions. The most peripheral bins are excluded from the analysis. Data point for centrality integrated yield ratio in $\sqrt{s_{NN}} =2.76$ TeV p-p collisions is unavailable.}
\end{table*}

	Note that, similar to Pb-Pb collision system, there is an indication of lowering of the $T_f$ for higher multiplicity events, though the value of $T_f$ for the high multiplicity class is almost~1.7 times larger than the $T_f$ we obtained for the centrality integrated Pb-Pb system. The most intriguing feature that resembles the feature of the Pb-Pb system is the simultaneous freeze out of all the states at higher multiplicity events. Our analysis results for the p-Pb system is shown in the lower panel of the Fig.\eqref{Fig:Upsilon_SSystem}. Multiplicity dependence of the $T_f$ for mid-rapidity CMS data at $\sqrt{s_{NN}}=5.02$ TeV~\cite{CMS:2013jsu} and multiplicity integrated rapidity dependence of $T_f$ using ALICE data at $\sqrt{s_{NN}}=8.16$ TeV~\cite{ALICE:2019qie} are presented in the lower-left and lower-right panels respectively. Like p-p collision system, lowering of the $T_f$ and simultaneous freeze-out of different bottomonia states with increasing track multiplicity has been observed. One may note that, the $T_f$ extracted at the backward rapidity is slightly higher than that of the forward-rapidity region, but it is not completely conclusive because of the large uncertainty. Implementation of finite phase space coverage of the data is not seen to generate any visible effect on the resulting $T_f$ values.

\begin{figure*}[t]
	\includegraphics[height=2.0in,width=3.3in]{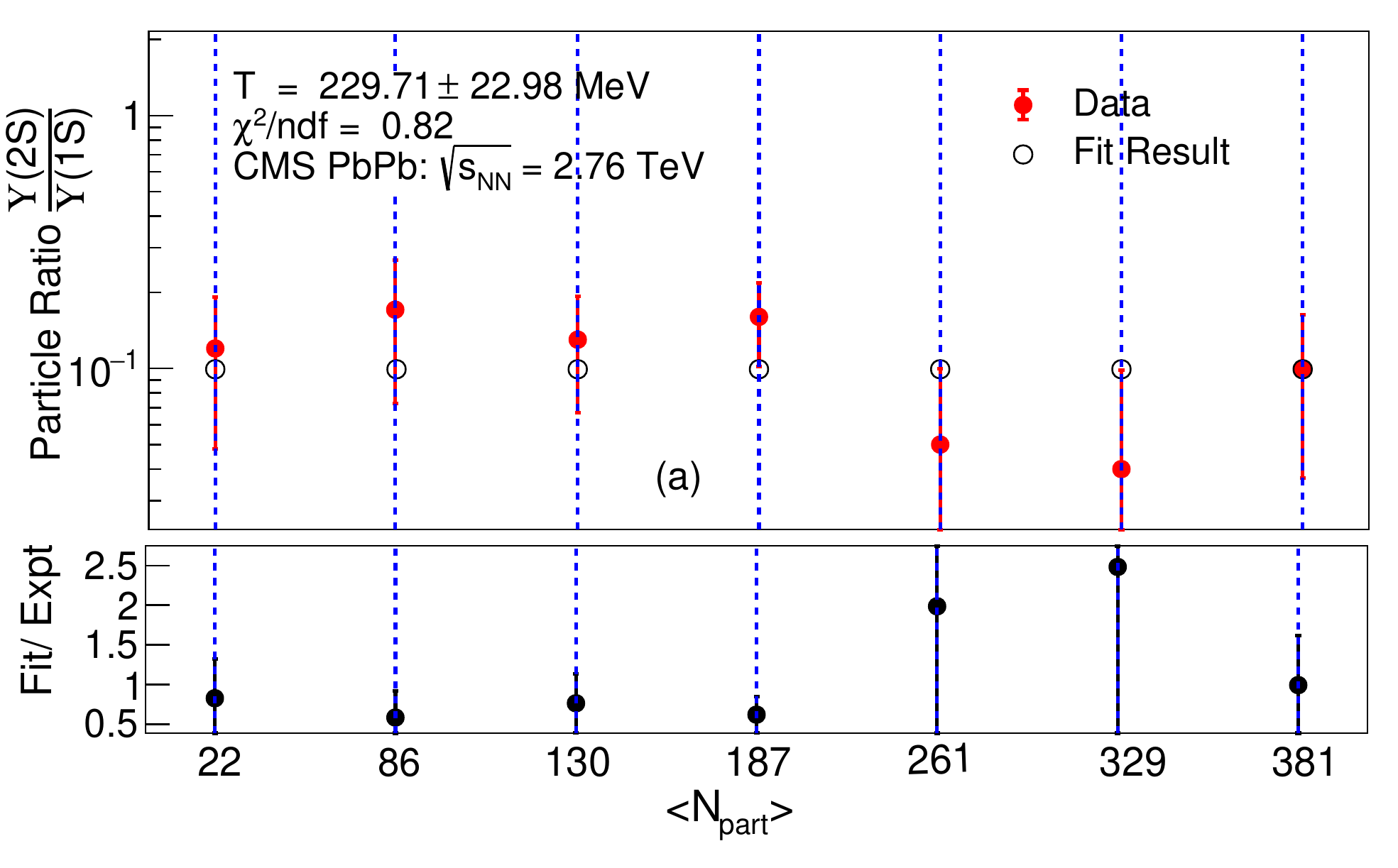}
	\includegraphics[height=2.0in,width=3.3in]{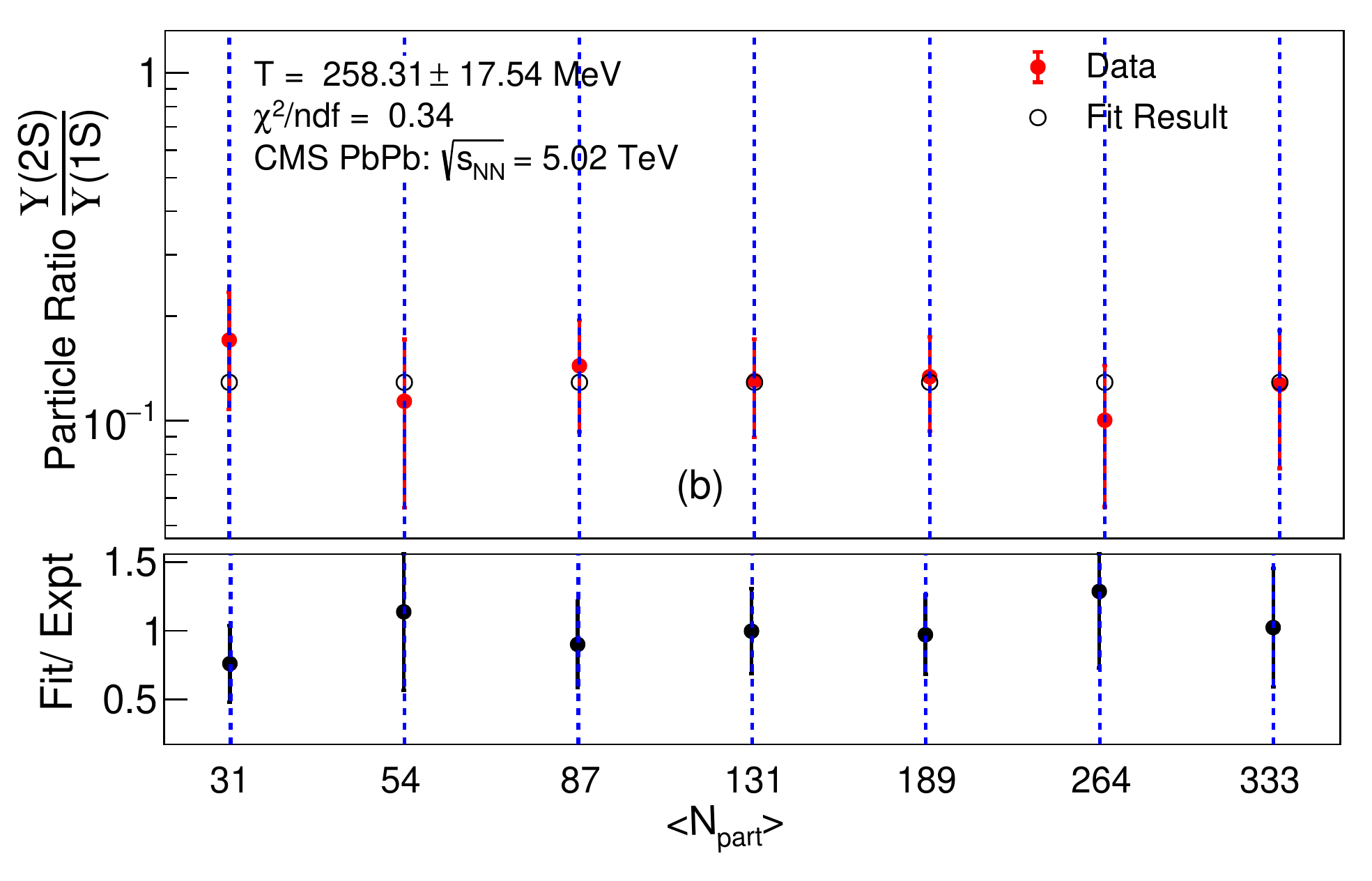}
	\includegraphics[height=2.0in,width=3.3in]{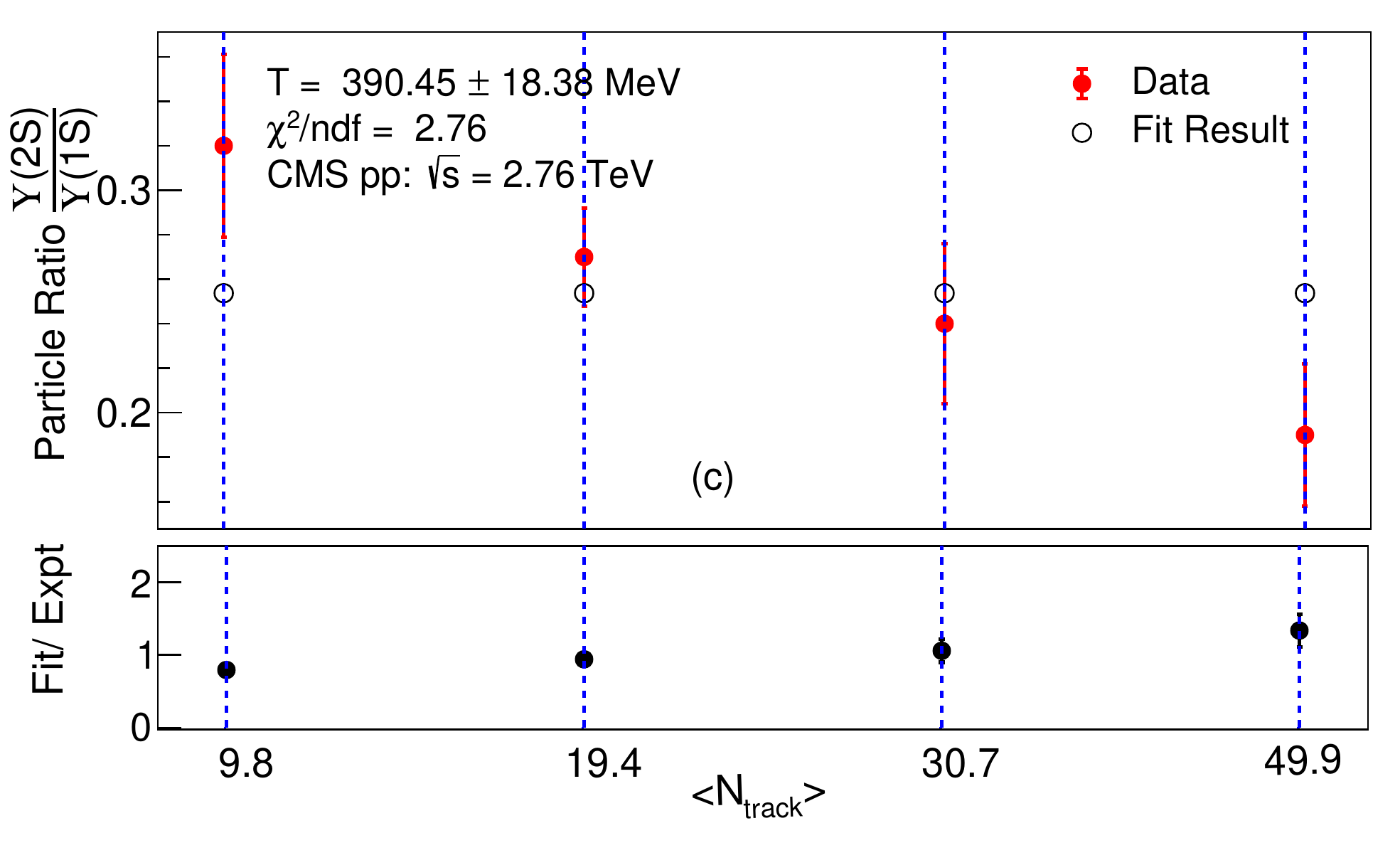}
	 \includegraphics[height=2.0in,width=3.3in]{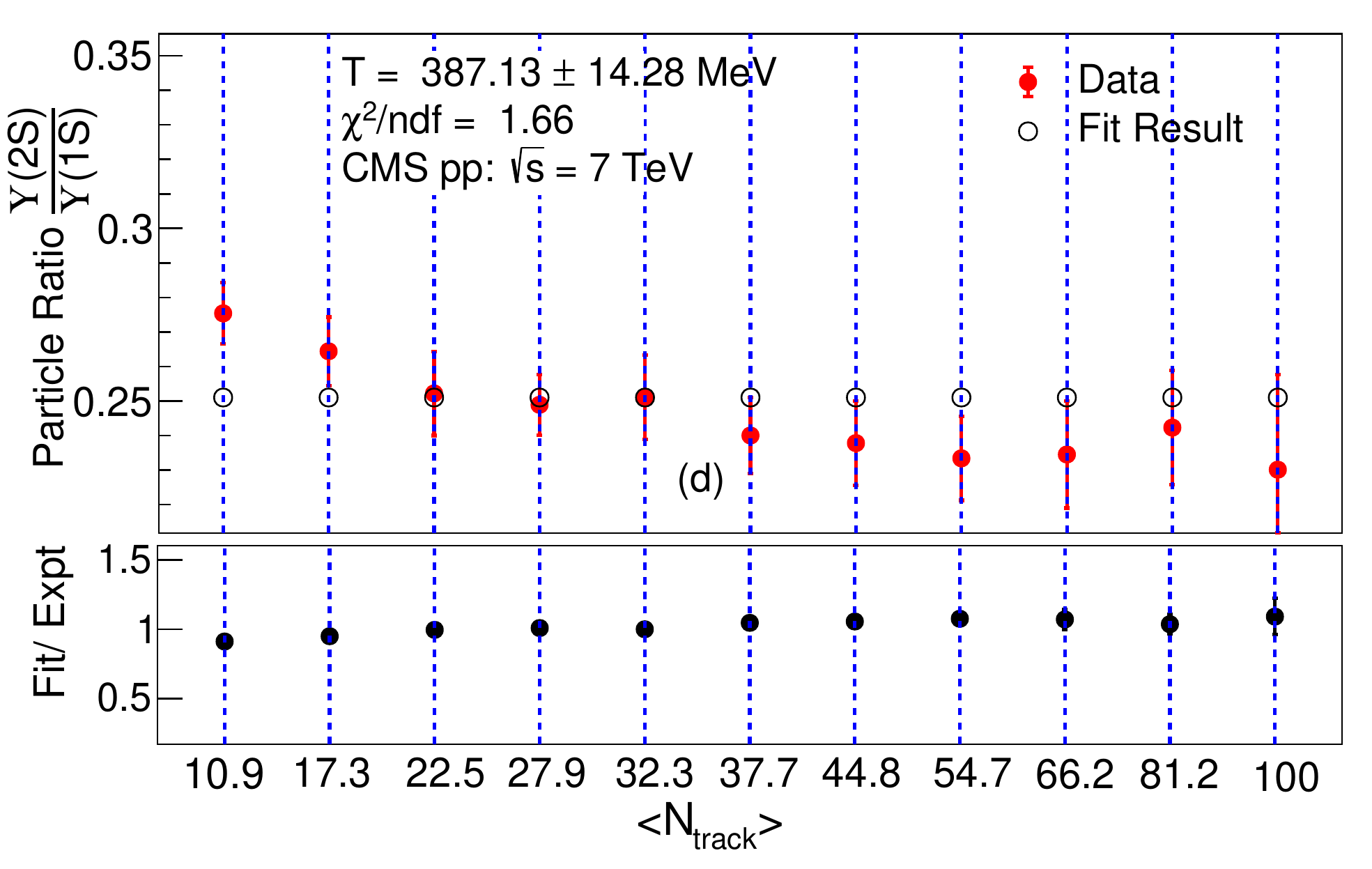}
	\caption{Single parameter ($T_{f}$) thermal model fit to the excited ($\Upsilon (2S)$) to ground state ($\Upsilon(1S)$) ratio of the bottomonia states for various centrality bins in (a) $\sqrt{s_{NN}} = 2.76$ TeV and (b) $\sqrt{s_{NN}} = 5.02$ TeV Pb-Pb and in (c) $\sqrt{s_{NN}} = 2.76$ TeV and (d) $\sqrt{s_{NN}} = 7$ TeV p-p collisions. The most peripheral bins are excluded from the analysis. The fit-to-data ratio are also shown for each system.}
	\label{fig:simul_fit}
\end{figure*}

	We have also analyzed the relative yields of charmonium states measured in p-Pb collisions at $\sqrt{s_{NN}}= 5.02$ TeV~\cite{ALICE:2016sdt} and $\sqrt{s_{NN}} = 8.16$ TeV \cite{ALICE:2020tsj}, and also in p-p collisions at $\sqrt{s_{NN}}=$ 5.02, 7, 8, 13 TeV as made available by the ALICE collaboration~\cite{ALICE:2021qlw,ALICE:2014uja,ALICE:2015pgg,ALICE:2017leg}. In this sector also, we analyzed the freeze-out temperatures considering both with and without the acceptance correction effect. Moreover, like all previous cases, the experimental $p_T$ range goes down to zero thus no effect of the acceptance correction has been found. To exclude the non-prompt contribution, the $f_B$-values in p-Pb system are taken as $12.4 \pm 0.89 \%$ at forward rapidity and $8.37 \pm 1.12 \%$ at backward rapidity at 5.02 TeV~\cite{LHCb:2013gmv}. At 8.16 TeV, the corresponding contribution is $14.51 \pm 1.06 (10.99 \pm 1.1) \%$ at forward (backward) region~\cite{LHCb:2017ygo}. Similarly, in p-p collisions, the $f_B$-values at $\sqrt{s} $ = 5.02, 7 and 13 TeV are taken as $15.7 \pm 2.3 \pm 1.5 \%$ (measured in kinematic domain $p_{T} > 1$ GeV/c, $|y| < $0.9), $~ 20 \%$ (since in the $p_{T}$ range $0-10$ GeV/c $f_B$ varies between $10 - 30 \%$) and $18.5 \pm 1.5 \pm 1.4 \%$ respectively as available in~\cite{ALICE:2021edd}. 

	Though the ATLAS collaboration measured the non-prompt contribution of the $J/\psi$ as a function of $p_T$ in the 8 TeV p-p collision, the starting $p_T$ range is as large as 8 GeV \cite{ATLAS:2015zdw}. Thus to analyze 8 TeV ALICE data~\cite{ALICE:2015pgg} in the $p_T$ coverage, $0<p_T<12$,  we consider the same variation of $f_B$ in the concerned $p_T$  range as has been found in others  $\sqrt{s} $, which are around 10 to 30\%. In our analyses the variation of $f_B$ with $p_T$ has accounted accordingly for all the cases. The finding of this analysis for p-Pb and p-p is presented in Fig.\eqref{fig:Jpsi_pPb} and \eqref{fig:Jpsi_pp} respectively. The freeze-out temperature as a function of collision centrality (expressed in terms of average number of binary collisions) is shown separately for forward (lower right panel) and backward (lower left panel) rapidity regions. No significant collision energy or rapidity dependence of $T_f$ has been observed. Apart from the most central and most peripheral data points (at 8.16 TeV), where the uncertainty is huge, for all other data points in both rapidity intervals, there is an overall tendency to decrease in $T_f$ with increasing collision centrality, but not so conclusive. As earlier, the extracted $T_f$ for all investigated data sets are significantly larger than the $T_f$ we obtain from analyzing the relative yield of the Pb-Pb system at 2.76 TeV ALICE data. (see. Fig.~\eqref{fig:Jpsi_PbPb}). The unnaturally high value of quarkonia freeze-out temperature obtained for small collision systems might point towards the fact that the produced quarkonia states are far from thermalized. Alternatively it might also occur that the life time of the produced fireball in these collisions is quite small leading to the early freeze-out of these quarkonia states.
	
        To further investigate the possibility of quarkonia thermalization in small collision systems at LHC, we adopt the following heuristic. We perform a single parameter ($T_{f}$) thermal model fit of the excited ($\Upsilon(2S)$) to ground state ($\Upsilon(1S)$) yield ratio of the bottomonia states for various centrality bins in Pb-Pb and p-p collisions. The most peripheral bins are excluded from the analysis. The fit results are displayed in Fig.~\eqref{fig:simul_fit} and summarized in Tab.~V. The resulting fit quality is not very different  Pb-Pb and p-p collisions. The $\chi^{2}$ per degree of freedom ($\chi^2$/ndf) values in the range $ 1 - 3$ for p-p collisions can be attributed to the large error bars associated with the corresponding yield ratio data points. Interestingly the extracted $T_{f}$ values in case Pb-Pb collisions is similar within errors to that obtained from the centrality integrated yield ratio, that plausibly indicates the thermal nature of these heavy hadrons in Pb-Pb collisions. However for p-p collisions, the two $T_{f}$ values (one extracted from the combined fit of yield ratios in  different centrality bins and the other calculated from the centrality integrated yield ratio) are largely different. This makes the thermal description of quarkonia in p-p collisions questionable. We refrain from performing this exercise for bottomonium production in p-Pb collisions due to small number of reported centrality bins.  Similarly, the charmonium sector is difficult as centrality bins are not reported for p-p collisions.

\section{Summary} 

In this article, we have analyzed the relative yields of different bottomonia and charmonia states produced in Pb-Pb collisions at LHC, within a thermal statistical model framework. The underlying assumption is the early thermalization and subsequent freeze-out of these heavy hadrons resulting in their chemical freeze-out at temperatures, significantly higher than that of light and strange hadrons. All three bottomonium states are seen to be simultaneously frozen in the fireball, at a temperature of around 230 MeV, independent of collision energy. No significant rapidity dependence of $T_f$ has been observed as a possible outcome of boost invariance. The systematic dependence of $T_f$ on the collision centrality is investigated in details. Insensitivity of the extracted $T_f$ values on the detector phase space coverage possibly indicates towards the thermalization of the $\Upsilon$ states as measured by different experimental collaborations. 
In the charmonia sector the data collected by CMS collaboration suffers from the presence of a very high value of low $p_T$ threshold, thus unsuitable for any thermal analysis. Analysis of the ALICE data suggests an early thermalization of the charmonia in 5.02 TeV Pb-Pb collisions. Our investigations are also extended to analyze the available quarkonium production in p-Pb and high multiplicity p-p collisions. Though the qualitative pattern of the freeze-out systematics in these small collision systems are similar to those seen in heavy-ion collisions, the  $T_f$ values appear to be unrealistically large. Further investigations are necessary to make any robust conclusion about the thermalization of heavy hadrons in small collision systems. 

\section{Acknowledgement}
The authors would like to thank Dr. Subikash Choudhury for many stimulating discussions during the peparation of the manuscript.


\bibliographystyle{unsrt}
\bibliography{Quarkonia_CFO_v11.bib}

\end{document}